\documentclass[fleqn,usenatbib]{mnras}

\usepackage{newtxtext,newtxmath}

\usepackage[T1]{fontenc}
\usepackage{ae,aecompl}
\usepackage[toc,page]{appendix}
\usepackage{enumitem}
\usepackage{natbib}
\usepackage{amsmath,bm}

\usepackage{graphicx}	
\usepackage{amsmath}	
\usepackage{tabularx}
\usepackage[dvipsnames]{xcolor}
\usepackage{upgreek}
\usepackage{ulem}
\usepackage{float}
\newcommand{\kms}{\mbox{$\>{\rm km\, s^{-1}}$}}
\newcommand{\pc}{\>{\rm pc}}
\newcommand{\kpc}{\mbox{$\>{\rm kpc}$}}
 
\newcommand{\kmsk}{\mbox{$\>{\rm kpc\, km\, s^{-1}}$}}
\newcommand{\Gyr}{\mbox{$\>{\rm Gyr}$}}
\newcommand{\Myr}{\mbox{$\>{\rm Myr}$}}

\newcommand{\Msun}{\>{\rm M_{\odot}}}

\newcommand\degrees{^\circ}
\newcommand\gaia{{\it Gaia}}

\newcommand{\avg}[1]{\mbox{$\left<{#1}\right>$}}

\newcommand{\ie}{{\it i.e.}}
\newcommand{\eg}{{\it e.g.}}

\newcommand{\thetajg}{\mbox{$\theta_{L,\,\mathrm{gas}}$}}

\newcommand{\thetaj}{\mbox{$\theta_{L}$}}

\defcitealias{wrinkle}{SD18}
\defcitealias{huang}{H18}

\title[Bending waves from warps]{Bending waves excited by irregular gas inflow along warps}

\author[]{Tigran Khachaturyants$^{1}$\thanks{E-mail: astrotkh@gmail.com},
	Leandro {Beraldo e Silva}$^{1,2}$, Victor P. Debattista$^{1}$, 
	Kathryne J. Daniel$^2$\\
	$^1$ Jeremiah Horrocks Institute, University of Central Lancashire, Preston, PR1 2HE, UK \\
	$^2$ Department of Astronomy, University of Michigan, 1085 S. University Ave., Ann Arbor, MI 48109, USA \\
	$^3$ Bryn Mawr College, Department of Physics, Bryn Mawr, PA 19010, USA \\
}

\date{Accepted XXX. Received YYY; in original form ZZZ}

\pubyear{2021}

\begin{document}
	\label{firstpage}
	\maketitle
	
	\begin{abstract}
    \gaia\ has revealed clear evidence of bending waves in the vertical kinematics of stars in the Solar Neighbourhood. We study bending waves in two simulations, one warped, with the warp due to misaligned gas inflow, and the other unwarped. We find slow, retrograde bending waves in both models, with the ones in the warped model having larger amplitudes. We also find fast, prograde bending waves. Prograde bending waves in the unwarped model are very weak, in agreement with the expectation that these waves should decay on short, $\sim$ crossing, timescales, due to strong winding. However, prograde bending waves are much stronger for the duration of the warped model, pointing to irregular gas inflow along the warp as a continuous source of excitation. We demonstrate that large amplitude bending waves that propagate through the Solar Neighbourhood give rise to a correlation between the mean vertical velocity and the angular momentum, with a slope consistent with that found by \gaia. The bending waves affect populations of all ages, but the sharpest features are found in the young populations, hinting that short wavelength waves are not supported by the older, kinematically hotter, populations. Our results demonstrate the importance of misaligned gas accretion as a recurrent source of vertical perturbations of disc galaxies, including in the Milky Way.
	\end{abstract}
	
	
	
	
	\section{Introduction}
	\label{sec:Introduction}
	
	Early hints that the Solar Neighbourhood is vertically perturbed \citep{Gomez+12,Widrow+12,Yanny+13,Williams+13,Carlin+13, Faure+14} have been dramatically confirmed with the {\it Gaia} data. Already by using just the Tycho-\gaia\ DR1 Astrometric Solution (TGAS) dataset \citep{gaia1, gaia}, \cite{wrinkle} (hereafter SD18) found a linear increase in the mean vertical velocity, $\avg{v_z}$, with the azimuthal velocity, $v_\phi$, angular momentum, $L_z$, and guiding radius, $R_g$, of stars in the Solar Neighbourhood (SN). Since the line of nodes of the Galactic warp is only $\sim 17.5\degrees$ ahead of the Sun \citep{chen}, this linear increase is potentially the warp's direct imprint on the local stellar kinematics. \citetalias{wrinkle} only used stars along narrow cones in the centre and anti-centre directions to obtain $v_z$ and $v_\phi$ without requiring radial velocity measurements, which the TGAS dataset lacks. \citetalias{wrinkle} also noted the presence of a wave-like pattern imprinted on the overall $\left<v_z\right>$ distribution, towards both the centre and anti-centre directions. A smooth, monotonic warp would not present such a signal; instead \citetalias{wrinkle} obtained a good fit with a simple sinusoidal wave, as might be produced by a winding warp or by a bending wave. Subsequently, this pattern in \avg{v_z}\ was replicated by \cite{huang} (hereafter H18) using a $\sim 10^{5}$-star sample from the LAMOST-TGAS data. Unlike the TGAS sample, LAMOST-TGAS has full 6D phase space measurements, which allowed \citetalias{huang} to replicate the wave-like pattern in \avg{v_z}\ versus $L_z$ and versus $v_{\phi}$ in the entire SN. With the release of \gaia\ DR2 the linear increase and wave-like pattern in $\avg{v_z}$ were again confirmed by \cite{friske+19}. 
	
	Subsequently, \gaia\ DR2 revealed a phase-space spiral in the $(z,v_z)$~plane. \cite{antoja} selected $\sim 9\times10^{5}$ stars in the Solar annulus of \gaia\ DR2 RVS, a sample containing the full 6D phase-space coordinates \citep{dr2rv}, and projected them onto the $(z,v_z)$~plane. The result was a spiral with one complete wrap, with a trailing tail reaching up to $\sim 700 \pc$ and $\sim 40\kms$ in $|z|$ and $|v_z|$, respectively. This phase-space spiral is particularly apparent when colour-coded by the azimuthal velocity, $v_\phi$, implying a coupling between the in-plane and vertical motions. The presence of this phase-space spiral indicates that the SN is undergoing vertical phase mixing as a result of vertical perturbations \citep{antoja}. The \textit{Gaia} phase-space spiral was dissected by \cite{juntai}, who showed it is weaker for stars on radially hotter orbits.
	
	The cause of these vertical perturbations in the SN remains uncertain. In general, vertical perturbations in galactic discs propagate as bending waves \citep{hunter_toomre, merritt_sellwood, sellwood1996, sellwood1998, Kazantzidis, cheque, khoper,darling, galah} with many possible causes. \cite{widrow2012} presented evidence for a wave-like perturbation in the Galactic disc in the form of the Galactic North-South asymmetry, speculating it could have come about via satellite interaction. \cite{feldman} used a high-resolution numerical simulation to study the interactions of dark matter substructure with the disc and observed that subhalo interactions resulted in distinct and coherent variations in the vertical velocities of disc stars. \cite{gomez} presented multiple high-resolution cosmological simulations of individual Milky Way-sized galaxies that develop significant vertical bending waves via satellite interactions.
	
	The Sagittarius dwarf galaxy (Sgr) is the most frequently invoked external cause of vertical excitation of the Milky Way's disc \citep{sgr_ibata, sgr_dehnen, laporte2019}, due to its relatively recent ($\sim 0.4 - 1$ Gyr ago) passage through the disc and an orbit that is perpendicular to the Galactic plane \citep{sgr_ibata, laporte2019}. Sgr has also been suggested to be the cause of the bending wave observed by \citetalias{wrinkle} and \citetalias{huang}.
	The analysis of the phase-space spiral led \cite{antoja} to infer that the Galactic disc was perturbed in the past $300 - 900\Myr$, which matches current estimates of a pericentric passage by Sgr. \cite{juntai} presented further support for the Sgr scenario with a vertically perturbed test particle simulation. They estimated that the perturbation should have happened at least $500\Myr$ ago to observe the \textit{Gaia} phase-space spiral in its current form. However, other simulations have shown inconsistencies in the Sgr scenario. \cite{binney} produced a phase-space spiral in a SN population extracted from a distribution function fitted to \gaia\ DR2 RVS and estimated that the spiral formed $400\pm150\Myr$ ago. While their time scale estimate is in some agreement with \cite{antoja}'s results, the mass and duration of the interaction required to produce similar phase-space spirals were significantly higher and faster, respectively. In a pure $N$-body simulation \cite{Hawthorn+21} have shown that the current mass estimate of the Sgr dwarf is too low to excite the phase spiral. Instead \cite{Hawthorn+21} suggested that the interaction had to have happened $1-2\Gyr$ ago with the Sgr dwarf losing mass at a high rate. Additionally, \cite{Bennett+21} used one-dimensional (vertical) models of satellite-disc interaction and were unable to reproduce the observed asymmetry in the vertical number counts for any plausible combination of Sgr and Milky Way properties.
	
	On the other hand, \citet{cheque} showed that isolated galaxies can also self-excite bending waves. Their $N$-body simulations of isolated galaxies naturally develop bending waves not just when the halo is clumpy but also when it is a smooth distribution of a finite number of dark matter particles. The bending waves in both kinds of simulations have a similar morphology and frequencies, but differ in amplitude, with the clumpy halo exciting waves of higher amplitudes. In the smooth halo models, the bending waves were seeded by the random noise of the halo and bulge particle distributions \citep{chequers+17}. In the clumpy halo models, instead, the subhalos imprint local perturbations on the disc which then shear into bending waves. Buckling of a galactic bar also induces bending waves in the disc. \cite{khoper} presented a high resolution $N$-body simulation that developed a bar which then buckled, \ie\ it suffered a vertical bending instability of the bar \citep{raha+1991, sellwood}. The resulting bending waves propagated outward in the disc and remained coherent for a long time, with the phase-space spirals still being distinguishable 3 Gyr after the bar buckled.  However it is unclear whether the Milky Way's bar could have buckled this recently without scattering too many relatively young stars into the bulge \citep{vpd2019}.
	
	An alternative mechanism for generating vertical bending waves comes from the observation that, in galaxy formation simulations, gas reaches the disc with a misaligned angular momentum \citep{binney_may86,ostriker_binney89,vdb2002, roskar2010, velliscig, stevens, earp+19}, regardless of whether it settles to the disc via hot or cold modes. Such angular momentum misalignments cause long-lived warps, as opposed to the transient warps excited by interactions \citep{ostriker_binney89, roskar2010, aumeraccr}. In the presence of a live dark halo, the assumption of steady warp precession \citep{Toomre83,Dekel&Shlosman83} fails as warps in $N$-body simulations rapidly wind up \citep{Binney_1998}. However, even in the presence of a live halo, misaligned accretion could still create warps with amplitudes comparable to that of observed warps \citep{Jiang&Binney99}.
	These warps provide another mechanism by which the disc may be vertically excited, as shown by \citet{gomez}, who found that some of their cosmological simulations had prominent vertical bends in discs with no recent satellite interaction. They argued that these bends are most prominent in the youngest stellar populations ($< 2 \Gyr$) and cold gas, and almost absent in the oldest stars. The Milky Way's H{\sc i} disc has long been known to be warped \citep{Kerr,weaver_williams,2006ApJ...643..881L}, with the warp reaching  $\geq4 \kpc$ above the midplane at $R=25\kpc$. A warp has also been observed in the stellar component of the Galactic disc \citep{efremov,reed,lopez}. Recently, with the help of the \textit{WISE} catalogue of periodic variable stars \citep{wise}, the stellar warp has also been mapped in greater detail in the young stellar populations \citep{chen}.
	
	This paper uses $N$-body+SPH (Smooth Particle Hydrodynamics) simulations to explore a scenario in which bending waves are induced by gas accreting along a warp.  In Paper {\sc I} \citep{Khachaturyants+21} we used the same warped simulation to show that if the MW's warp is formed by misaligned gas accretion, then stars formed in the warp could migrate inwards and be found in the SN. The paper is organised as follows: we describe a warped and a control unwarped simulation in Section~\ref{sec:Simulations}. The evolution of the warp is described in Section~\ref{sec:warpevol}. In Section~\ref{sec:bending_waves} we analyse the bending waves that develop in both the warped and unwarped simulations, comparing and contrasting them. Lastly we summarise our results in Section~\ref{sec:Conclusions}.

	\section{Simulations}
	\label{sec:Simulations}
	
	We construct two simulations, one with, and one without a warp, in order to study the effect of warps on generating bending waves. Hereafter we refer to the simulations as the warped and unwarped models.
	
	\subsection{The warped model simulation}
	The warped model has the same initial conditions as the simulation used in \cite{Khachaturyants+21} and is produced via the method of \citet{vpd2015}, which constructs triaxial dark matter models with gas angular momentum misaligned with the principal axes of the halo.
	\citet{AumerWhite2013} showed that inserting rotating gas coronae within non-spherical dark matter halos leads to a rapid and substantial loss of gas angular momentum.  To avoid this catastrophic angular momentum loss, our approach includes adiabatic gas already while merging haloes to produce a non-spherical system.  We merge two identical spherical Navarro–Frenk–White (NFW) \citep{NFW1996} dark matter haloes, each having a co-spatial gas corona comprising 10 per cent of the total mass.

	The mass and virial radius of each dark matter halo at $z=0$ is set to $M_{200} = 8.7 \times 10^{11}\Msun$ and $r_{200} = 196 \kpc$, respectively. The gas is in pressure equilibrium within the global potential. Gas velocities are initialised to give a spin parameter of $\lambda = 0.16$ \citep{Bullock+2001}, with specific angular momentum $j \propto R$, where $R$ is the cylindrical radius. Both the dark matter halo and the gas corona are comprised of $10^6$ particles. Gas particles start out with masses of $1.4 \times 10^5 \Msun$ and softening $\epsilon = 20 \pc$, while dark matter particles have two different mass configurations ($10^6 \Msun$ and $3.6 \times 10^6 \Msun$ inside and outside $200 \kpc$, respectively) and $\epsilon = 100 \pc$.  The two halos are placed $500 \kpc$ apart and approach each other head-on at $100 \kms$.  If the direction of the separation vector (and the relative velocity) is the $x$-axis and the halos are rotating about their $z$-axes, a tilt about the $y$-axis is applied to one of the halos so that the final system will be prolate with its long axis along the $x$-axis and a gas angular momentum tilted with respect to the symmetry axes of the halo.
	
	This simulation is evolved with the smooth particle hydrodynamics code {\sc gasoline} \citep{Wadsley+2004}, with a base time-step $\Delta t = 10 \Myr$. This time-step is refined for individual particles such that each particle satisfies the condition $\delta t = \Delta t/2^n < \eta \sqrt{\epsilon/a_g}$, where $a_g$ is the acceleration at the particle's current position, with $\eta = 0.175$. The opening angle of the tree code calculation is set to $\theta = 0.7$.
	
	The result of this setup is a dark matter halo with $r_{200} = 238 \kpc$ and $M_{200} = 1.6\times10^{12}\Msun$, and gas with $\lambda = 0.11$.  These are the initial conditions of the warped simulation. At this stage we turn on gas cooling, star formation and stellar feedback using the blastwave prescriptions of \citet{Stinson+2006}. Gas particles form stars with a $10\%$ efficiency if a gas particle has number density $n > 1$ cm$^{-3}$, temperature $T <15,000$ K and is part of a convergent flow.  
	
	Star particles form with an initial mass that is 1/3 of the initial parent gas particles, corresponding to $4.6 \times 10^4\Msun$ at our resolution. The star particles all have $\epsilon = 20 \pc$.  Once a gas particle loses $80\%$ of its initial mass, the remaining mass is distributed amongst the nearest neighbouring gas particles, leading to a decreasing number of gas particles. Star particles are represented by an entire stellar population with a Miller–Scalo \citep{MillerScalo1979} initial mass function. The evolution of star particles includes asymptotic giant branch stellar winds and feedback from Type II and Type Ia supernovae, with their energy injected into the interstellar medium (ISM). Each supernova releases $10^{50}$ erg into the ISM. The time-steps of gas particles satisfy the additional condition $\delta t_{\rm gas} = h\, \eta_{\rm courant}/[(1 + \alpha)c + \beta\, \mu_{\rm max}]$, where $h$ is the SPH smoothing length, $\eta_{\rm courant} = 0.4$, $\alpha = 1$ is the shear coefficient, $\beta = 2$ is the viscosity coefficient, $c$ is the sound speed, and $\mu_{\rm max}$ is the maximum viscous force measured between the gas particles \citep{Wadsley+2004,Springel10}. The SPH kernel uses the 32 nearest neighbours. Gas metallicity is taken into account in the gas cooling process using the prescriptions of \citet{Shen+2010}; to prevent the cooling from dropping below our resolution, we set a pressure floor on gas particles of $p_{\rm floor} = 3G\epsilon^2\rho^2$, where $G$ is Newton’s gravitational constant, and $\rho$ is the gas particle’s density \citep{Agertz+2009}.

	\subsection{The unwarped simulation}	
	
	The unwarped model is the M1\_c\_b simulation described in \cite{karl21+}. Briefly, the model is similar to one of the spherical models we start with in the warped simulation, except that the initial gas angular spin is $\lambda = 0.065$ \citep{Bullock+2001}.  Feedback via supernova explosions again employs the blastwave prescription \citep{Stinson+2006}.  The main difference between this and the warped simulation (aside from the initial conditions) is that we use a gas particle softening of $50 \pc$, the star formation efficiency is $5\%$ and the feedback from supernovae is set to $4\times10^{50}$ erg per supernova.

	\subsection{Pre-processing the simulations}
	\label{ssec:preproc}
	
	Simulation snapshots are saved every $10 \Myr$ and are processed through our custom \textsc{Python} library suite \citep{Khachaturyants+21}. The processing involves centring the galactic disc and then rotating it into the $(x,y)$ plane based on the angular momentum of the inner stellar disc ($R< 5\kpc$) for both  models.
	After this reorientation, the warped model is rotated such that the maximum vertical displacement of the tilted ring model \citep{briggs}, \ie\ the warp's major axis (WMA), is on the $x$-axis and, consequently, the line of nodes (LON) is on the $y$-axis. Lastly, the disc is rotated by $180\degrees$ about the $y$-axis, which results in a sense of rotation (clockwise when viewed from the positive $z$-axis, hereafter the North Galactic pole) and warp orientation similar to that of the Milky Way \citep{chen}. We define an azimuthal angle coordinate $\phi_w$, where $\phi_w=0$ represents the ascending node of the LON ($y<0$ axis), and increases in the direction of rotation. As a result, the gas warp in each snapshot reaches its peak negative value along the positive $x$-axis, \ie\ $\phi_w = -90\degrees$. In the case of the unwarped model the process is repeated without the WMA reorientation, so we define $\phi=0$ as being along the $x$-axis.
	
	\begin{figure}
		\includegraphics[width=.9\linewidth]{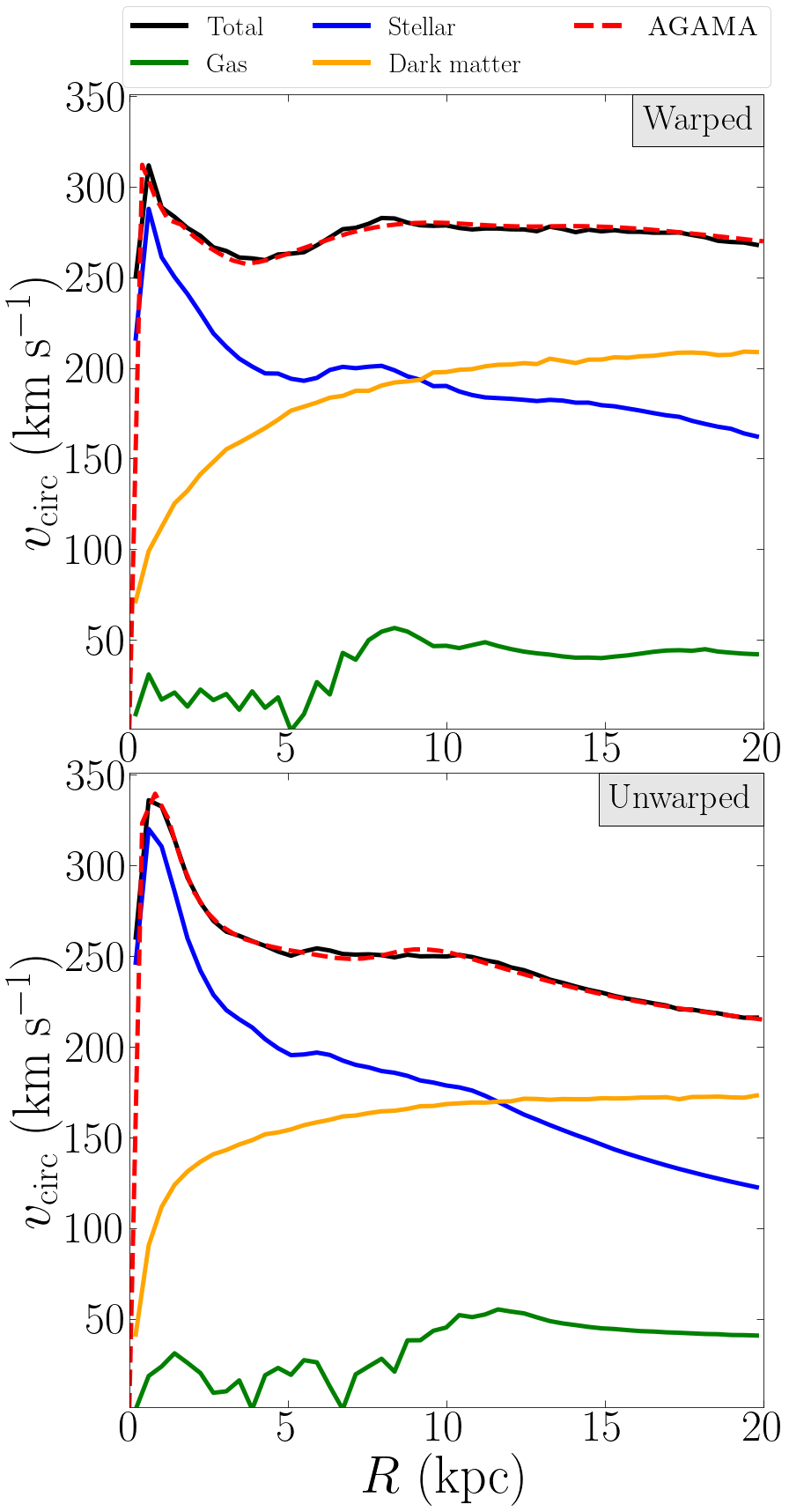}
		\caption{Rotation curves for the warped (top) and unwarped (bottom) models at $12 \Gyr$. Solid lines represent the rotation curves of each galactic component \citep[computed for each family of particles using the \textsc{profile} function of the \textsc{pynbody} library,][]{pynbody}, while the interpolated total potential \citep[computed with AGAMA,][]{agama} is represented by the dashed red lines.}
		\label{fig:rotcurve_ws_uw}
	\end{figure}
	
	The rotation curves of the two models at $12 \Gyr$ are presented in Fig.~\ref{fig:rotcurve_ws_uw}. Potentials for both simulations were interpolated using the {\sc agama} software library \citep{agama} using a single multipole approximation for the stellar, gas, and dark particles combined. Rotation curves of the interpolated potentials are presented in Fig.~\ref{fig:rotcurve_ws_uw} as dashed red lines. As in the Milky Way, the rotation curves of the two models are relatively flat, though the unwarped model has a higher stellar density in the centre and therefore a peak in the rotation curve at $\sim 1\kpc$. 
	
	\section{Warp evolution}
	\label{sec:warpevol}
	\begin{figure*}
	\centerline{
	    \includegraphics[width=1\linewidth]{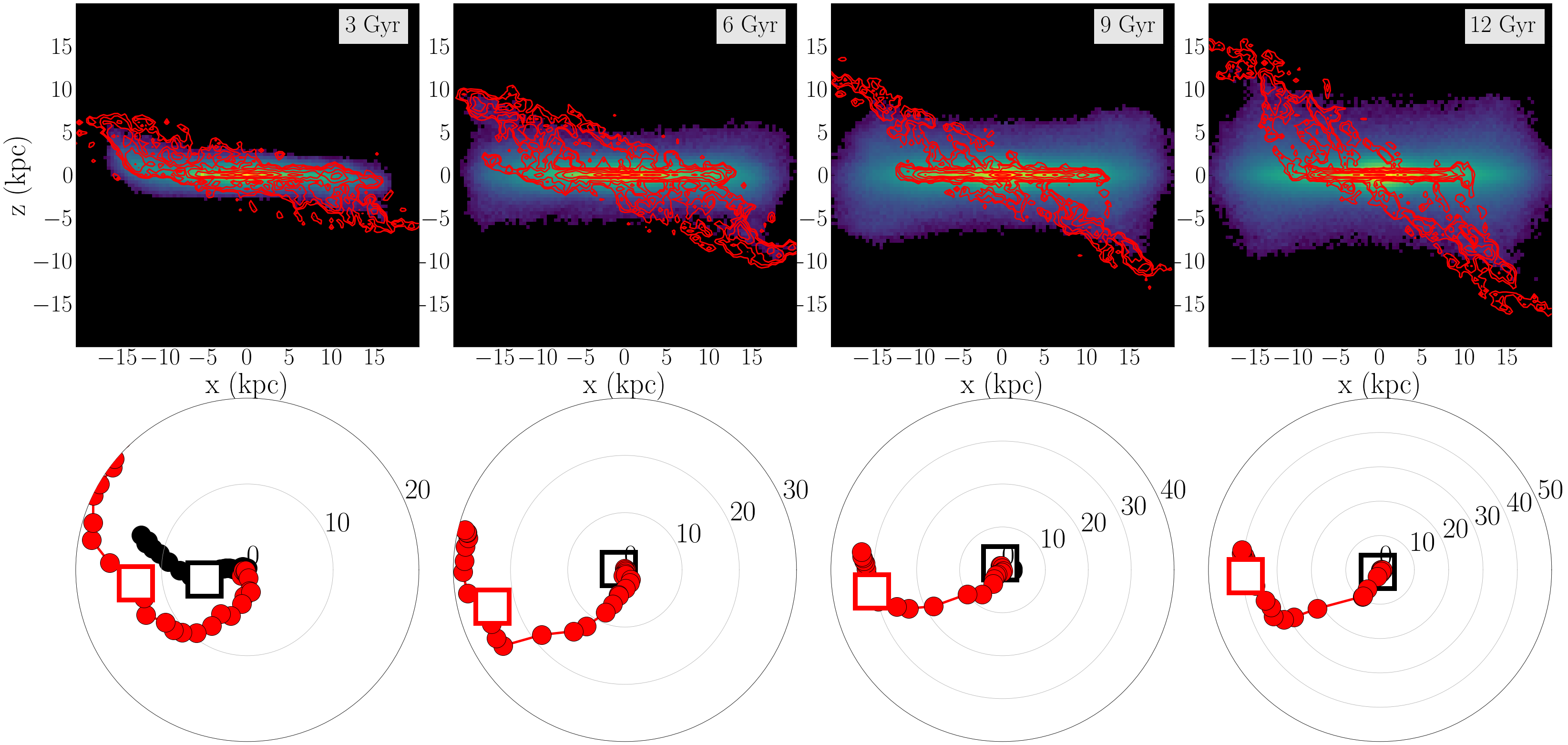}}
		\caption{Top row: edge-on views of the stellar and cold gas ($T_g\leq50,000$ K) distributions at four times in the evolution of the warped model. The colour represents the stellar surface density, while the red contours represent the cold gas column density. The times are labelled at the top-right in each panel. A warp is present throughout the evolution of the warped model. The simulation is rotated so that the major axis of the warp is along the $x$-axis. The warp reaches heights  $|z|\sim15\kpc$ over this evolution. Bottom row: Briggs figures for the warped model showing the evolution of the stellar (black) and cool gas (red) warps at the same times. Markers represent annuli with $\Delta R = 0.5\kpc$, equally spaced from $5$ to $20\kpc$, with the square markers indicating $R=15\kpc$. Annuli containing a total mass that is $\leq 10^6\Msun$ are not shown. The stellar disc is somewhat warped at $t=3\Gyr$ but becomes flatter throughout its evolution.}
		\label{fig:gas_edgeon_ws}.
		\centerline{\includegraphics[width=1\linewidth]{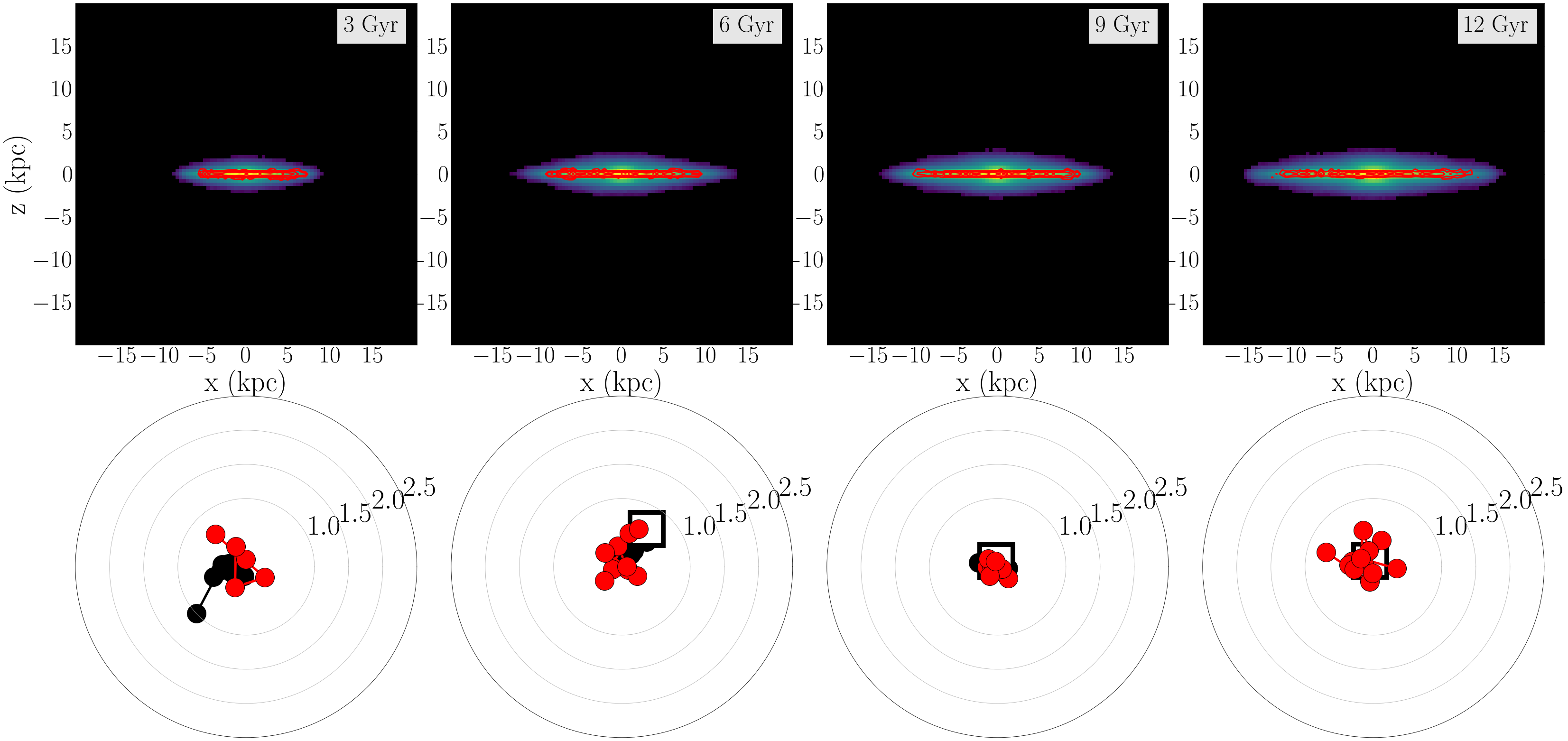}}
		\caption{Similar to Fig.~\ref{fig:gas_edgeon_ws} but for the unwarped model. In contrast to the warped model, there are no warps in either the gaseous or stellar components in the edge-on distributions. The Briggs figures have a reduced scale with max $\theta=2.5\degrees$ set as the upper limit, so even though we see some changes at different radii, both gas and stellar discs are quite flat throughout the model's evolution.}
		\label{fig:gas_edgeon_uw}.
	\end{figure*}

	The top row of Fig.~\ref{fig:gas_edgeon_ws} presents edge-on views of stars (colour) and the cool ($T_g\leq50,000$ K) gas (red contours) between $t=3\Gyr$ and $12\Gyr$. Throughout the evolution of the warped model, gas is accreting onto the disc along an integral-shaped warp. By $12\Gyr$ the gas warp extends up to $15\kpc$ above the plane at $R \sim 20\kpc$. Because of our re-orientation of the disc, the major axis of the warp is along the $x$-axis and reaches a peak negative value along the $x>0$ side; in reality, viewed from an inertial frame the disc is tilting slowly and continuously during this time \citep{binney_may86, ostriker_binney89, vpd2015, earp+17, earp+19} but we subtract this tilting. 
	
	In order to study the evolution of the warp, we construct Briggs figures \citep{briggs} for the warped and unwarped models.  A Briggs figure represents warping by means of the spherical azimuthal and inclination angles, $\phi_{J}$ and $\theta_{J}$ respectively, between the total angular momenta of concentric annuli and the $z$-axis.  These are then plotted as the radial, $\rho$ (for $\theta_J$), and angular, $\psi$, (for $\phi_J$) variables of a 2D polar plot. Because we reorient the discs into the $(x,y)$ plane based on the angular momentum of the inner disc stars before we perform any analysis, the inner disc is at the origin of the Briggs figures, \ie\ it has angular momentum along the $z$-axis. 
	The bottom row of Fig.~\ref{fig:gas_edgeon_ws} shows Briggs figures for the warped model at the same time intervals. The figure presents the stars (in black) and the cool gas (in red) separately. The stellar and gaseous discs are divided into annuli of width $\Delta R = 0.5 \kpc$, and then we calculate the total angular momentum of particles in each annulus. A warp is present in the gas component throughout the evolution of the warped model. The warp grows slowly with time; by $12\Gyr$ it extends to almost $40\degrees$.  The warp traces a leading spiral relative to the sense of rotation of the disc, in agreement with Briggs's third rule of warp behaviour \citep{briggs} which states that, beyond a certain radius, the line of nodes lies along a loosely wound, leading spiral. On the other hand, the stellar component loses its large-scale warp after $6\Gyr$, and only a small stellar warp remains.

	In contrast, a similar analysis on the unwarped model does not reveal any notable disc warping. In the top row of Fig.~\ref{fig:gas_edgeon_uw} the edge-on views of the unwarped model present no stellar (colour) or gaseous (red contours) warps at any point in time. The bottom row of Fig.~\ref{fig:gas_edgeon_uw} shows similar Briggs figures as in Fig.~\ref{fig:gas_edgeon_ws} but with significantly smaller $\theta_L$ upper limits to underline the lack of warping in the unwarped model. We observe no warping in the stellar component at all times and only minor tilting at $R=10\kpc$ at $6$ and $9 \Gyr$ for the gas.
	
	\begin{figure}
		\includegraphics[width=.9\linewidth]{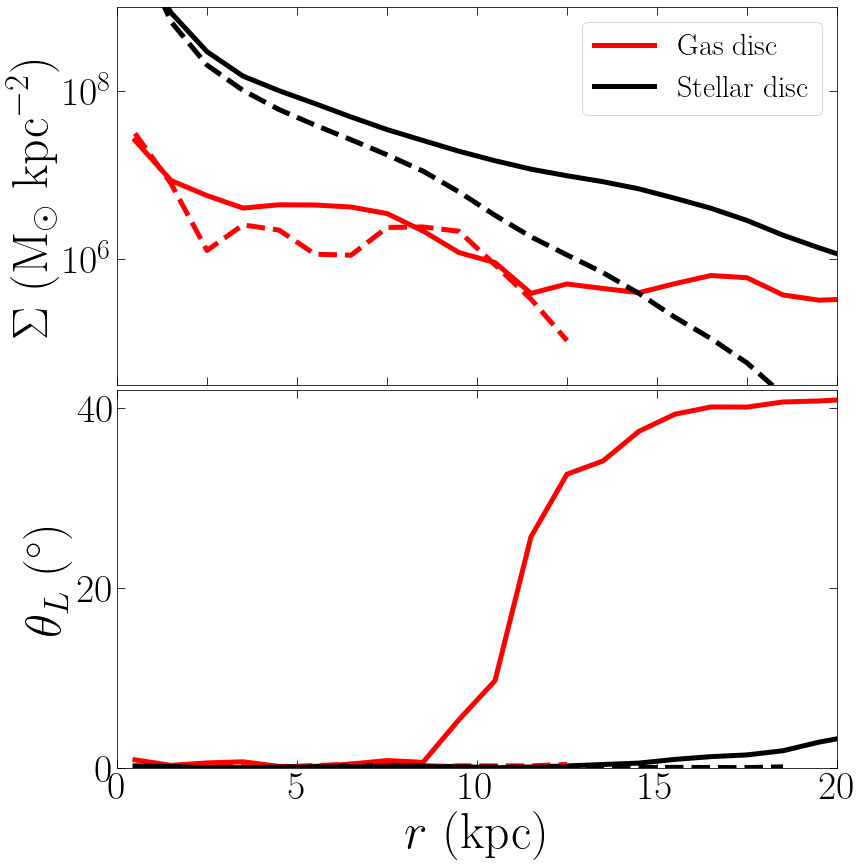}
		\caption{Profiles of the surface density, $\Sigma$, (top) and $\thetajg$ (bottom) in the warped (solid lines) and unwarped (dashed lines) models at $t=12\Gyr$. The profiles are shown for both the cold gas (red) and stellar (black) discs. Only bins containing a total mass $\geq7\times10^6 \Msun$ are shown.}
		\label{fig:briggs_theta}
	\end{figure}		
	
	Fig.~\ref{fig:briggs_theta} presents cold gas and stellar profiles for both models at $t=12\Gyr$ in the surface density (top panel) and their inclination (bottom panel). The unwarped model exhibits a drop in the gas surface density at the edge ($10 \kpc$) while the inclination of both stellar and gas components  remains flat throughout the disc and its outskirts. In the warped model we observe a slower decline in the surface density of both components, with the stellar disc showing a weak increase in inclination, caused by newly ($\leq 2\Gyr$) formed warp stars. The inclination of the gas disc rapidly grows from $r\geq 9 \kpc$ and reaches $\thetaj\sim40\degrees$ by $r\geq 15 \kpc$.

	\section{Bending waves}
	\label{sec:bending_waves}
	\subsection{The presence of vertical bends}
	\begin{figure*}
		\includegraphics[width=1.\linewidth]{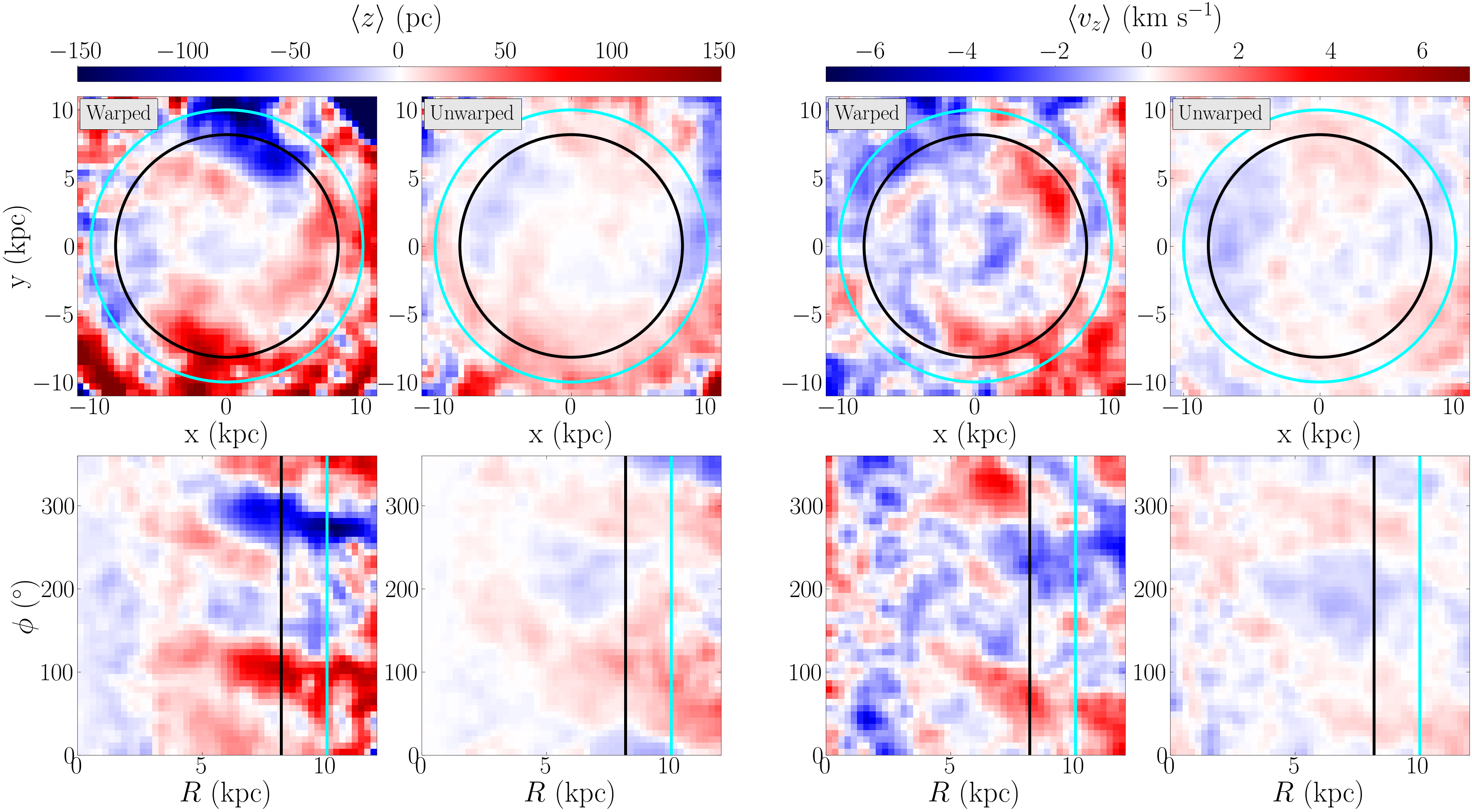}
		\caption{Distributions of the stellar mean height, $\avg{z}$ (left), and mean vertical velocity, $\avg{v_z}$ (right), for the warped and unwarped simulations (see top left annotation) at $t=11.7\Gyr$ in the $x-y$ (top) and $R-\phi_w$ (bottom) planes. The sense of rotation is clockwise (top row) and towards increasing $\phi$ (bottom row). A Gaussian filter has been applied to the colour distribution in each panel with a standard deviation of the Gaussian kernel set to $\sigma=1$ pixel $=450\pc \times 450 \pc$ (top) and $\sigma=1$ pixel $=250  \pc \times 7.5 \degrees$ (bottom). The solid black and cyan circles (vertical lines) represent the Solar annulus, $R=8.18\kpc$, and $R=10\kpc$, in the $x-y$ ($R-\phi$) plane, respectively.}
		\label{fig:z_ws_uw}
	\end{figure*}
	
	Fig.~\ref{fig:z_ws_uw} shows the stellar distributions in the $x-y$ (top) and $R-\phi_w$ (bottom) planes of the average height, $\avg{z}$ (left), and the average vertical velocity, $\avg{v_z}$ (right), in the warped and unwarped models (see top left annotation). The distributions highlight the presence of large-scale bends (coherent blue and red structures) in the disc. The warped model exhibits bends with amplitudes of $\sim 100 \pc$ and $\sim 3\kms$ for $\avg{z}$ and $\avg{v_z}$, respectively, that reach far inside the disc down to $R\simeq 2\kpc$. In agreement with \citet{cheque}, bends are also observed in the unwarped model, but are noticeably weaker, with amplitudes of $\sim25 \pc$ and $\sim 1 \kms$, respectively. The bends in the unwarped model also reach far inside the disc reaching the very centre. At first sight, the structure of the bends in both models do not appear to have any distinct shape and wavelength, requiring a more in-depth spectral analysis of the surface $\avg{z}$ distributions to probe for bending waves.
    
    \subsection{Spectral analysis of bending waves}
    \label{sec:spectral}
    Bending waves propagating in a kinematically cool galactic disc can be seen as the superposition of a "fast" (+) and "slow" (-) waves circulating with frequencies $\omega = m\Omega(R) \pm \nu(R)$, where $\Omega(R)$ is the angular rotation curve, $\nu(R)$ is the vertical frequency, and $m$-fold rotational symmetry is assumed. Inclusion of the disc's self-gravity raises $\nu(R)$, making bending waves more stable, \ie\ stiffer (contrary to density waves) -- see \cite{binney_tremaine_08}. The detailed inclusion of the disc's self-gravity, however, defies simple theoretical modelling, since it depends on the mass distribution of the bending wave itself. Moreover, the halo strongly reacts to the perturbed disc \citep[][]{Binney_1998}, making it impossible to model an equation of motion for the disc alone.

	We are thus left with the conservative constraint that ignores self-gravity and, for a given rotation curve $\Omega(R)$, bending waves can only propagate in regions that satisfy the condition
	\begin{equation}
	\label{eq:forbid_bend}
	 m^2\left[ \Omega_\mathrm{p} - \Omega(R)\right]^2 \geq \nu_h^2,   
	\end{equation}
	where $\Omega_\mathrm{p}= \omega/m$ is the pattern speed and $\nu_h$ is the frequency of vertical oscillation contributed by the halo potential. This defines, for $m=1$, a ``forbidden" region, $\Omega - \nu_h < \Omega_p < \Omega + \nu_h$, where bending waves cannot propagate \citep[e.g.][]{nelson_tremaine_1995}. 
	
	In flattened potentials, $\nu > \Omega$, so the "fast" wave is prograde, with a frequency $\omega$ depending strongly on $R$ for most radii, so differential rotation winds it up rapidly and it decays. The "slow" wave, on the other hand, is retrograde and circulates with frequency only weakly depending on $R$ for most radii. This wave is thus expected to wind up slowly and be long-lived.
	
	In order to investigate in detail the propagation of bending waves in our models, in this section, we employ the spectral analysis technique of \cite{sellwood_athanassoula_1986}, using a code based on that of \cite{Roskar+12}. This allows us to recover the spatial distribution and temporal evolution of pattern speeds. The code is applied to both the unwarped and the warped simulations, first for the density distribution and then for the vertical distribution.
	
	At each snapshot, we start by selecting star particles in concentric annuli. In each annulus, we first expand the azimuthal angular dependence of the normalised mass distribution in a Fourier series
	\begin{equation}
	    \mu(R, \phi) = 1 + \sum_{m=1}^\infty c_m(R)e^{-im\phi},
	\end{equation}
	with
	\begin{equation}
	    c_m(R) = \frac{1}{M(R)}\sum_{p=1}^N m_p e^{im\phi_p},
	    \label{eq:c_m}
	\end{equation}
	where the sum runs over particles inside the annulus, $m_p$ and $\phi_p$ are the mass and azimuth of particle $p$, respectively, and $M(R)$ is the total mass within the annulus. We calculate the coefficients $c_m(R)$ for every snapshot in a given time interval (hereafter baseline) and then perform a discrete Fourier transform of this time series as
	\begin{equation}
	    C_{m,k}(R) = \sum_{j=0}^{S-1}c_m(R,t_j)w_j e^{2\upi ijk/S},
	    \label{eq:C_mk}
	\end{equation}
	with $k=-S/2,...,S/2$, where $S$ is the number of snapshots in the baseline. The associated frequencies are given by
	\begin{equation}
	    \Omega_k = \frac{2\upi}{m}\frac{k}{S\Delta t},
	    \label{eq:Omega_k}
	\end{equation}
	where $\Delta t$ is the time between snapshots, and we adopt the Gaussian window function
	\begin{equation}
	    w(j) = e^{-(j-S/2)^2/(S/4)^2}.
	\end{equation}
	Finally, the power spectrum is computed as
	\begin{equation}
	    P(R,\Omega_k) = \frac{1}{W}|C_{m,k}(R)|^2,
	    \label{eq:power}
	\end{equation}
	where 
	\begin{equation}
	    W=S\sum_{j=0}^{S-1} w_j^2.
	\end{equation}
	We perform this calculation for a time baseline ${S\Delta t=1\,\Gyr}$, resulting in a resolution ${\Delta \Omega =2\upi/m\,\kms\kpc^{-1}}$ -- see Eq. \ref{eq:Omega_k}. We repeat this calculation for several time baselines, and the resulting power spectrum for the unwarped simulation is shown in Fig.~\ref{fig:739HF_hc_psp}.
	
	In order to analyse the bending signal, similarly to Eq. \ref{eq:c_m} we define
	\begin{equation}
	    \gamma_m(R) = \frac{1}{M(R)}\sum_{p=1}^N z_p m_p e^{im\phi_p},
	    \label{eq:gamma_m}
	\end{equation}
	where $z_p$ is the vertical height of particle $p$, and use Eqs. \eqref{eq:C_mk}-\eqref{eq:power} {\it mutatis mutandis}. Note that now $\gamma_m(R)$ is given in kpc and the associated power spectrum is given in $\kpc^2$.
	
	Finally, after calculating the power spectra, Eq.\eqref{eq:power}, for both density and bending signals, we identify the pattern speeds $\Omega_p$ as peaks in the radially-integrated power spectra, which we refer to as total power. However, while the disc surface density decreases exponentially with radius, the Fourier coefficients, Eqs.\eqref{eq:c_m} and \eqref{eq:gamma_m}, are normalised by the annulus total mass $M(R)$, giving "equal weights" to power at small or large radii. Thus, to better appreciate the relevance of different pattern speeds to the disc dynamics, the total power is weighted by the annulus mass:
	\begin{equation}
	    \mathrm{total\,power} (\Omega_k)=  \frac{\sum M^2(R)P(R,\Omega_k)}{\sum M^2(R)},
	    \label{eq:total_power}
	\end{equation}
	with the sum running over radial bins.  This total power is shown as curves next to the spectrograms in Fig.~\ref{fig:739HF_hc_psp}.
	
	For the analysis in this section, we also compute the frequencies of circular motion $\Omega(R)$ and radial oscillation $\kappa(R)$ produced by the total potential and the frequency of vertical oscillation produced by the halo $\nu_h(R)$, using {\sc agama} \citep{agama}. These frequencies are computed in the middle of each 1 Gyr baseline.
	
	\subsubsection{Unwarped simulation}
	
\begin{figure*}
		\includegraphics[width=1.\textwidth]{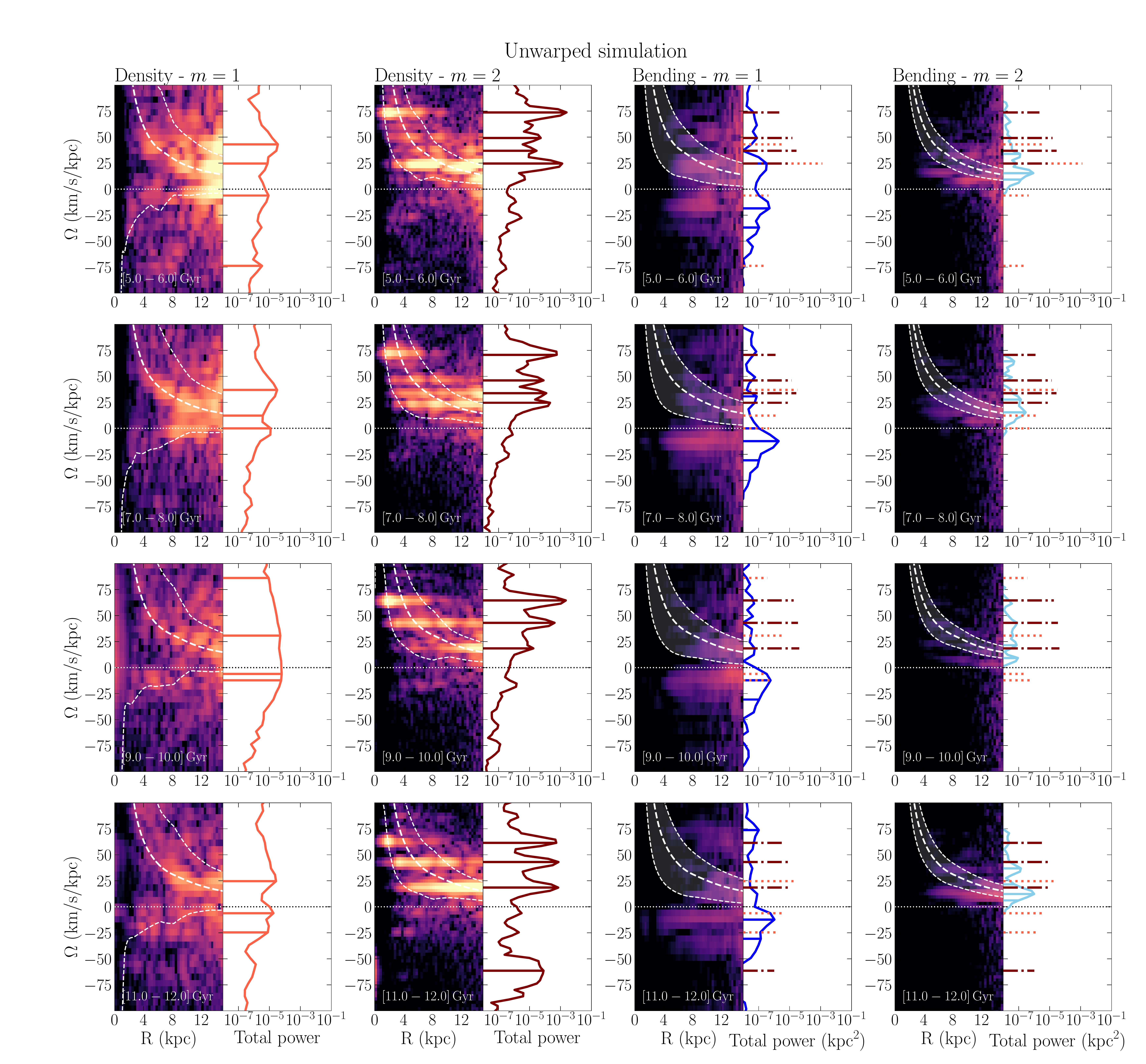}
		\caption{
		Power spectra for perturbations in the unwarped simulation at several time-intervals (rows). The first two columns show the power spectra for $m=1$ and $m=2$ density (bar$+$spiral) perturbations, with the mass-weighted radius-integrated power shown to the right of the spectrograms -- see Eq.\eqref{eq:total_power}. The thick and thin white dashed lines show $\Omega(R)$ and $\Omega\pm \kappa/m$, respectively. The two right-hand columns show the power spectra for $m=1$ and $m=2$ bending perturbations. The thick and thin white dashed lines show $\Omega(R)$ and $\Omega \pm \nu_h/m$, and the white shaded areas between these curves represent the forbidden regions for bending waves. For $m=1$, the expected long-lived slow retrograde motion is clearly visible, while the fast prograde pattern is weak. The total power peaks of the $m=1$ (light red dotted lines) and $m=2$ (dark red dot-dashed lines) density perturbations are repeated in the two right-hand columns.
		}
		\label{fig:739HF_hc_psp}
	\end{figure*}

	The two left-hand columns of Fig.~\ref{fig:739HF_hc_psp} show, for the unwarped simulation, the power spectra obtained for $m=1$ and $m=2$ density perturbations (as indicated in the titles) in the ($\Omega,R$) plane at different times (rows), from 5 Gyr to 12 Gyr. The $m=2$ density signal shows multiple pattern speeds at all times, covering a large radial extent and revealing the presence of multiple spiral density waves. The thick dashed white lines show the rotation curves, $\Omega(R)$, while the thin dashed white lines represent $\Omega \pm \kappa/m$.
	
	The panels at the right of the spectrograms show the (mass-weighted) total power (light and dark red), Eq.~\eqref{eq:total_power} (on a log scale), whose peaks reveal the pattern speeds; prominent peaks for $m=2$ are immediately distinguished. In an iterative scheme similar to that of \cite{Roskar+12} we identify the most prominent peak, fit a Gaussian function to it and subtract this Gaussian contribution from the total power. Then, we identify the next most prominent peak and repeat the process, identifying pattern speeds and power in the interval ${-100 \leq \Omega/\kms\kpc^{-1} \leq 100}$ up to a maximum of four peaks (horizontal lines, with length representing the power after the Gaussian subtraction of peaks previously identified). The $m=1$ density signal shows some significant power, but the peaks are not as prominent as those for $m=2$. It is interesting to note a prominent $m=2$ retrograde peak at the final baseline (bottom row), with power in the very inner disc. We verified that this is associated with a tiny bar which must have a prograde rotation so fast that the algorithm misinterprets it as a retrograde motion, given the simulation cadence. This peak is enhanced by the mass-weighted normalisation of the total power, Eq.~\eqref{eq:total_power}, but it is not of interest for our results. 

\begin{figure*}
	\includegraphics[width=1.\textwidth]{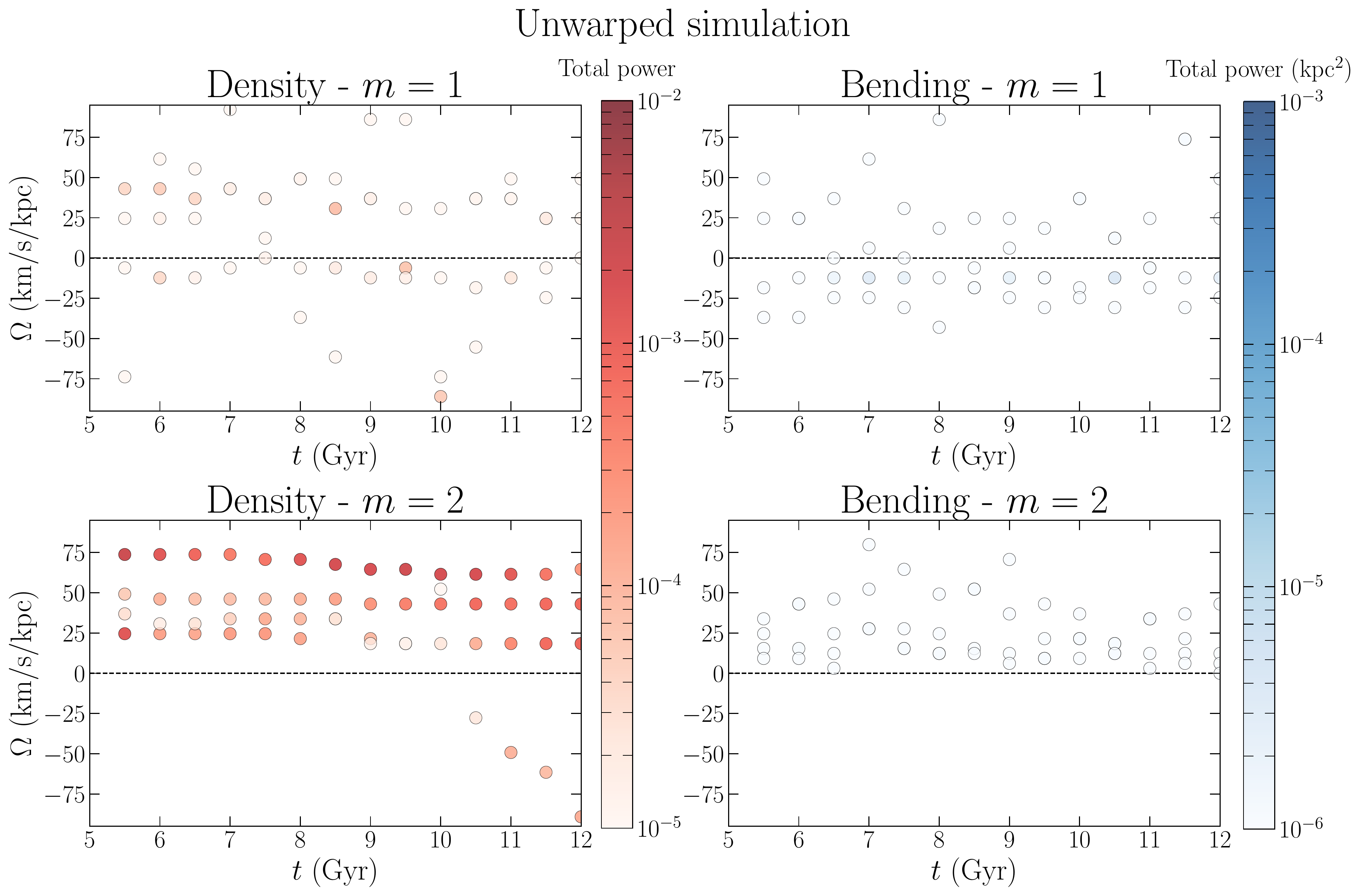}
	\caption{Pattern speeds in the unwarped model identified in Fig. \ref{fig:739HF_hc_psp}, for $m=1$ (top) and $m=2$ (bottom) density (left) and bending signals (right). Colours indicate the mass-weighted radially-integrated power. Different colour scales are chosen to differentiate between the density and bending pattern speeds. For the $m=2$ density signal, the gradual decrease of the highest pattern speed (upper points) is suggestive of a slowing bar (see centre-left column of Fig.~\ref{fig:739HF_hc_psp}). The other two discernible patterns are associated with spiral density waves, with the most prominent ones at $\Omega \approx 20-25\kms\kpc^{-1}$ and $\Omega \approx 40\kms\kpc^{-1}$. The $m=1$ bending plot shows the long-lived presence of a slow retrograde pattern (at $-15 \leq \Omega/\kms\kpc^{-1} \leq -10$) and (at some times) a very weak prograde signal at $\Omega \approx 24\kms\kpc^{-1}$.}
		\label{fig:739HF_hc_bubble}
	\end{figure*}

	Fig.~\ref{fig:739HF_hc_bubble} (left panels) shows the identified $m=1$ (top) and $m=2$ (bottom) density pattern speeds for several 1 Gyr-baselines, with colours representing the total power for a given pattern speed (Fig. \ref{fig:739HF_hc_psp}). Focusing on $m=2$, this figure clearly shows the simultaneous presence of multiple pattern speeds. The higher pattern speeds, at $60-75\kms\kpc^{-1}$, decreasing in time are due to the presence of a slowing bar \citep[see][]{karl21+}. The other two prominent pattern speeds can be attributed to the propagation of spiral density waves. In this simulation, for the time interval analysed, the pattern speeds show some evolution, changing values and power amplitude, but not very vigorously transient behaviour. The most prominent patterns are at $\Omega \approx 20-25 \kms\kpc^{-1}$ and $\Omega \approx 40 \kms\kpc^{-1}$. As shown in \citet{Roskar+12}, what is transient about these spirals is not necessarily their frequency, with some values seemingly preferred, but their amplitude, which continuously varies. We cannot exclude the possibility that certain modes are continuously being re-excited, most likely with random relative phases.
	
	The power spectra for the $m=1$ and $m=2$ bending signal in the unwarped simulation are shown in the two right-hand columns of Fig.~\ref{fig:739HF_hc_psp} (see the titles). The panels to the right of these spectrograms again show the (mass-weighted) radially-integrated power spectra (dark and light blue) with the peaks identified in the same way as before. The white thick dashed lines again show the rotation curve $\Omega(R)$, while the shaded white areas between $\Omega \pm \nu_h/m$ represent the forbidden regions for bending waves -- Eq.~\eqref{eq:forbid_bend}. Focusing on $m=1$, the most noticeable feature in these spectra is the ubiquitous presence of a slow retrograde pattern at ${-15 \lesssim \Omega/\kms\kpc^{-1} \lesssim -10}$ and extending inwards to $R\approx 5\kpc$, where $\Omega - \nu_h$ (bottom thin dashed curves) start to strongly depend on $R$ and severe winding is expected for kinematic bending waves.
	
	The right-hand panels of Fig.~\ref{fig:739HF_hc_bubble} show the evolution of the pattern speeds identified for the bending signals of $m=1$ (top) and $m=2$ (bottom) multiplicity. Focusing again on $m=1$, we confirm the ubiquitous presence of the slow retrograde mode, while prograde bending waves are barely noticeable. This seems in accordance with the theoretical expectation that, no matter how the bending perturbation is produced, the associated slow retrograde wave is long-lived, while prograde waves, if present, decay quickly. Interestingly, a prograde $m=1$ bending pattern, at $\Omega \approx 20-25 \kms\kpc^{-1}$ is detected at some snapshots, located inside the forbidden region for bending waves, but with very small power (see Fig.~\ref{fig:739HF_hc_psp}).

	\subsubsection{Warped simulation}
	
\begin{figure*}
		\includegraphics[width=1.\textwidth]{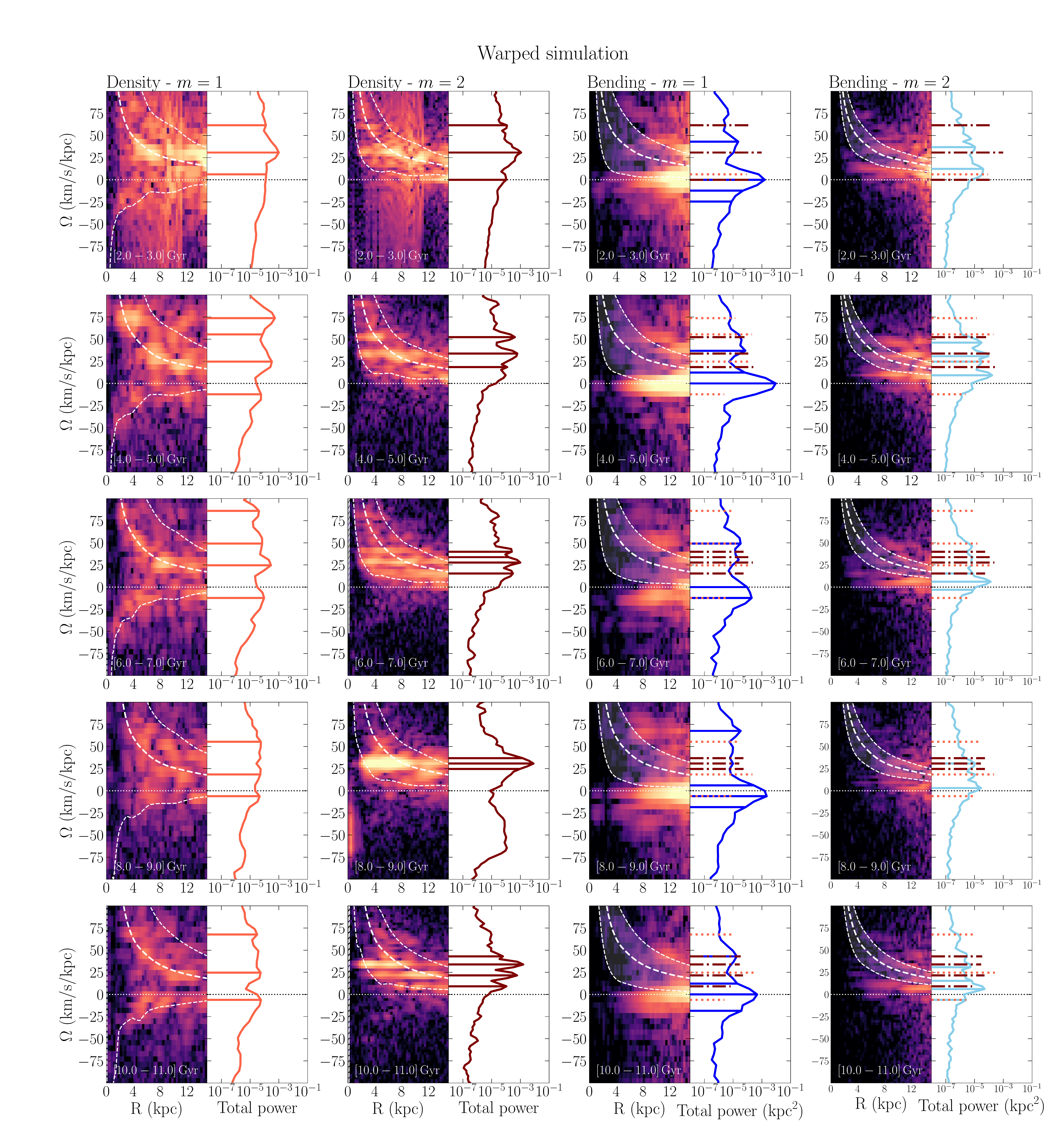}
		\caption{Similar to Fig. \ref{fig:739HF_hc_psp}, showing the power spectra at different times (rows) for the $m=1$ and $m=2$ density (left) and bending (right) perturbations in the warped simulation. The $m=2$  density panels show the simultaneous presence of various pattern speeds between the Lindblad resonances. In the $m=1$ bending panels, the most noticeable difference with respect to Fig. \ref{fig:739HF_hc_psp} is the strong peak at ${\Omega \approx 0\kms\kpc^{-1}}$, which is a trivial manifestation of the warp. As in the unwarped simulation, a slow retrograde motion is detected in the $m=1$ bending plot. Significant $m=1$ bending power is present for large $\Omega$ at large radii, \ie\ a fast prograde motion avoiding the forbidden region for bending waves, and peaking at $25 \leq \Omega/\kms\kpc^{-1} \leq 50$.} 
		\label{fig:670h_psp}
	\end{figure*}

\begin{figure*}
		\includegraphics[width=1.\linewidth]{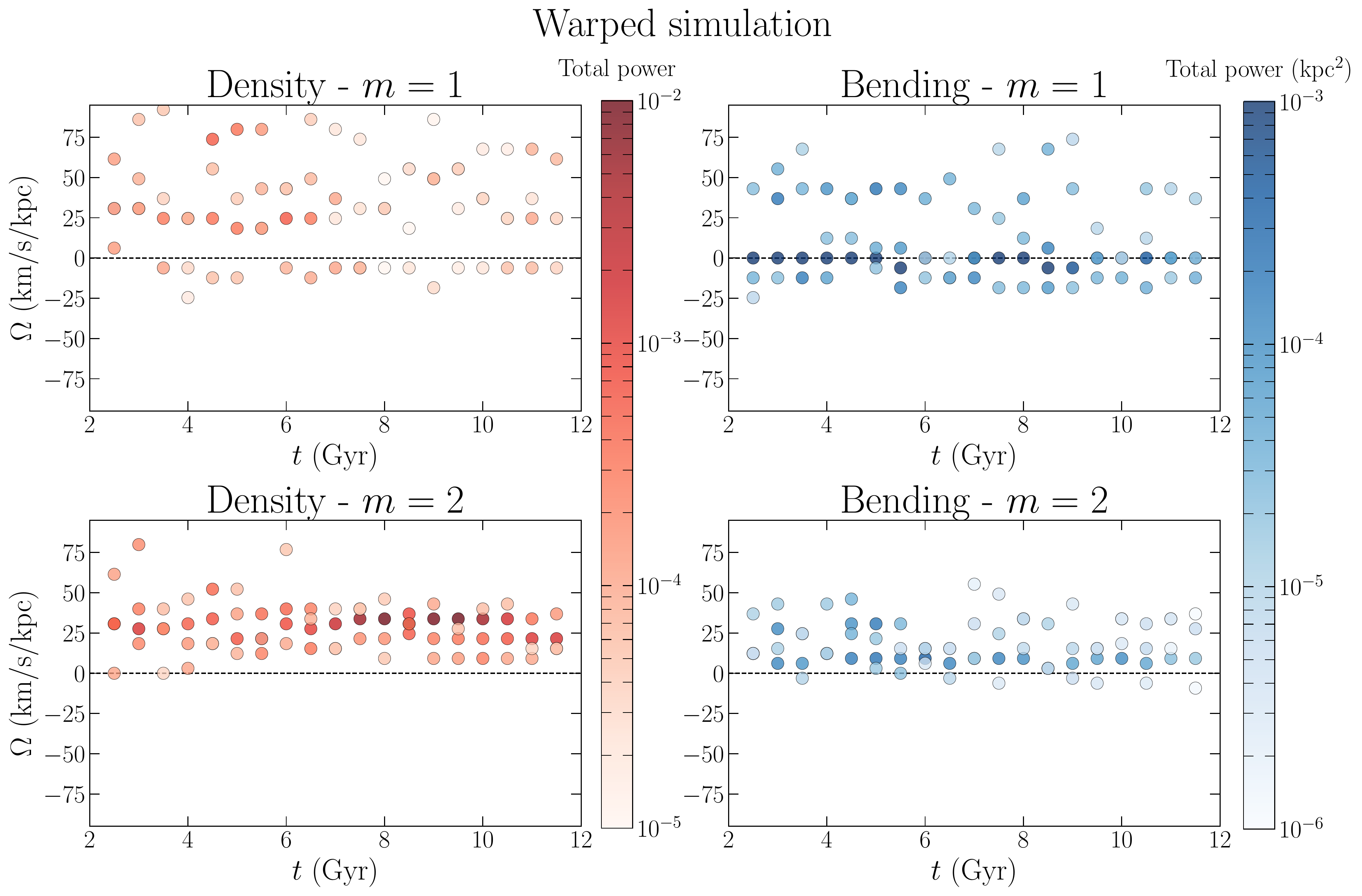}
		\caption{Evolution of the pattern speeds identified in Fig. \ref{fig:670h_psp}, for the warped simulation. The $m=2$ density panel reveals a more transient evolution, in comparison to the unwarped simulation. In the $m=1$ bending panel, the darkest points (power $\gg10^{-4}\kpc^2$) lying at $\Omega\approx 0\kms\kpc^{-1}$ represent a trivial manifestation of the warp. As in the unwarped simulation (Fig. \ref{fig:739HF_hc_bubble}), a persistent slow retrograde $m=1$ bending signal is detected. Unlike the unwarped simulation, a fast $m=1$ bending prograde motion ($25\leq\Omega/\kms\kpc^{-1}\leq 50$) is detected with substantial power.}
		\label{fig:670h_bubble}
	\end{figure*}

	Fig.~\ref{fig:670h_psp} shows the power spectra for the warped simulation (in this simulation we stored outputs at high cadence already from 2 Gyr, which permits us to perform spectral analysis from this point onwards), with the same scheme of density and bending $m=1$ and $m=2$ signals as in Fig.~\ref{fig:739HF_hc_psp}. As in the unwarped model, the $m=2$ density signal exhibits multiple pattern speeds present simultaneously, covering a large radial extent. The pattern speeds are not as sharply defined as in the unwarped simulation, which might be due to the perturbation from the warp. Alternatively, this could be due to the warped model being thicker: at $12 \Gyr$ and between $5\leq R/\kpc\leq10$ the discs of the warped and unwarped models have root-mean-square $z$, of $0.94\kpc$ and $0.55\kpc$, respectively.
	The left-hand panels of Fig.~\ref{fig:670h_bubble} show the time evolution of the pattern speeds identified in Fig.~\ref{fig:670h_psp} for the $m=1$ (top) and $m=2$ (bottom) density signal. Note that no bar forms in this simulation.
	
	The right hand panels of Fig.~\ref{fig:670h_psp} show the spectrograms for the bending signals. The bending $m=1$ waves exhibit a prominent peak at $\Omega = 0\kms\kpc^{-1}$ for almost all snapshots, which is the trivial signal of the warp itself. This peak is so prominent that it can visually hide nearby peaks, which our iterative peak finding and Gaussian subtraction scheme allows us to detect (dark blue horizontal lines). The $m=1$ peak due to the slow retrograde motion ($\Omega \approx -15\kms\kpc^{-1}$) is detected at almost all time-intervals. Additionally, significant power in fast prograde waves is now observed, at large radii, and peaking at $25 \leq \Omega/\kms\kpc^{-1} \leq 50$.
    
	The time evolution of the pattern speeds identified in Fig. \ref{fig:670h_psp} for the $m=1$ (top) and $m=2$ (bottom) bending signals are presented in the right-hand panels of Fig.~\ref{fig:670h_bubble}. As in the unwarped simulation (Fig. \ref{fig:739HF_hc_bubble}), we see the ubiquitous presence of a slow, retrograde $m=1$ wave in the warped simulation, with substantially more power than in that model.
	
	The main difference between the warped and the unwarped models is the presence of a strong, fast prograde motion in the $m=1$ bending signal, at ${25 \lesssim \Omega/\kms\kpc^{-1} \lesssim 50}$ in the warped system. These fast prograde patterns peak at large radii (see Fig.~\ref{fig:670h_psp}), thus avoiding the forbidden region for bending waves. Note that the prograde bending waves are present at all times. This is due to the long-lived nature of the warp in this simulation. While fast prograde bending waves are expected to decay quickly, the warp continuously perturbs the disc, re-exciting these waves.
	
	The main conclusion from the analysis in this section is that slow retrograde bending waves are present in both the unwarped and the warped models, throughout their evolution. On the other hand, only in the warped model are significant fast prograde bending waves detected, which must be persistently re-excited by the warp.
    
    \subsection{The source of the vertical perturbations}
	\begin{figure}
		\includegraphics[width=.9\linewidth]{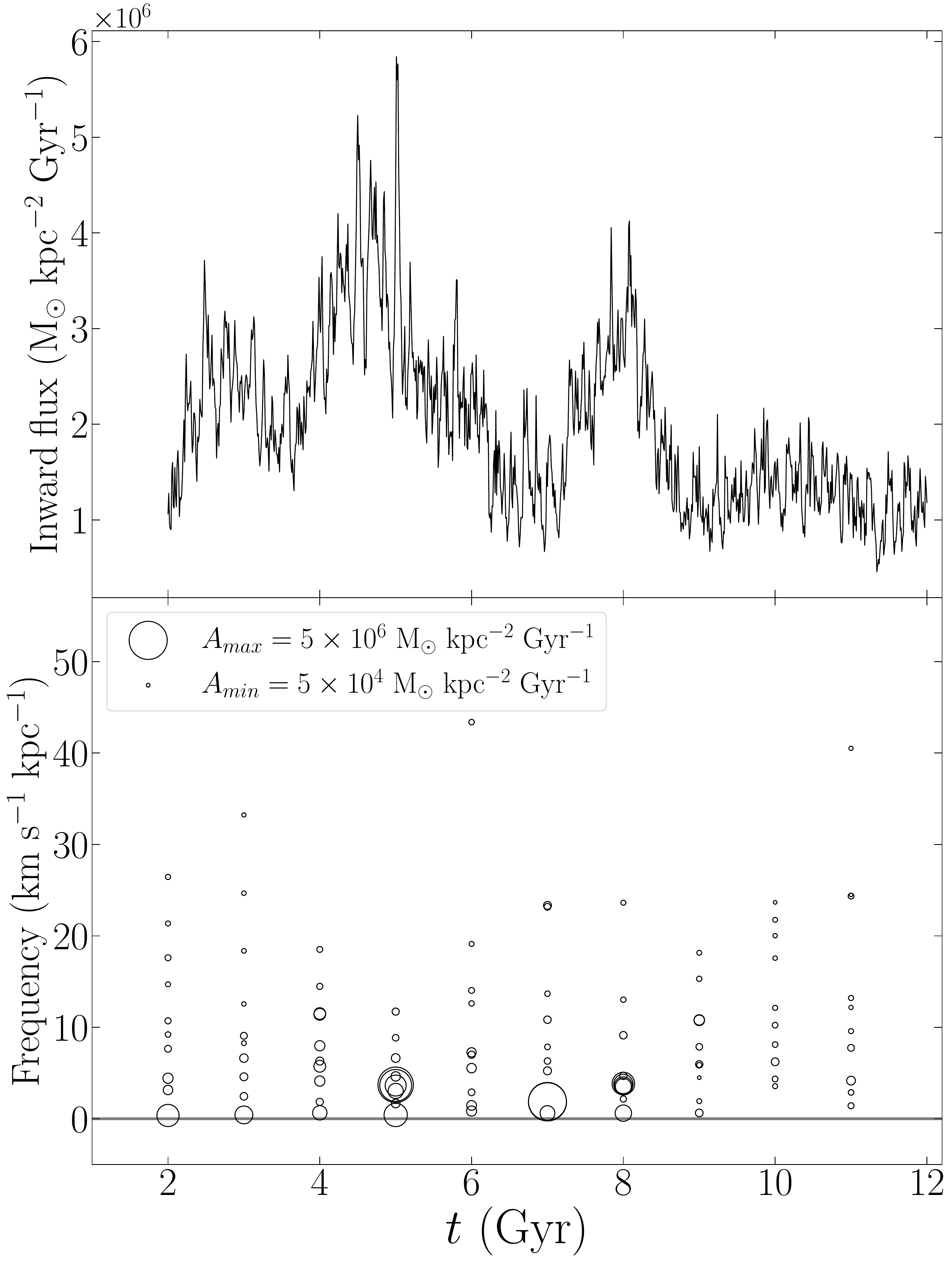}
		\caption{Top: evolution of the inward mass flux of cold gas ($\rm{T}<50,000$K)  through a spherical shell with $R=15 \kpc$ and $\delta R=0.2 \kpc$. Bottom: frequencies derived from a discrete Fourier transform of the mass flux on $1\Gyr$ baselines. The marker size indicates the amplitude with the values of the maximum and minimum amplitudes and their respective marker sizes shown in the legend.
		}
		\label{fig:bubble}
	\end{figure}
	In the previous section we demonstrated the presence of bending waves in both the warped and unwarped simulations, with different properties. Bending waves in simulations of unwarped isolated galaxies were already reported by  \cite{chequers+17}, who suggested shot noise in the dark matter halo as a source mechanism. Fast prograde bending waves are expected to dissipate rapidly and therefore be weak, as indeed we find in the unwarped simulation. However, in the warped model the consistent power in the prograde bending waves indicates that the disc in the warped model is continuously being vertically excited.
	
We now verify that the gas inflow along the warp is the source of these vertical perturbations. We start from the continuity equation,
\begin{equation}
\frac{\partial \rho}{\partial t} + \nabla \cdot(\rho \textbf{u}) = 0,
\end{equation}
where $\rho$ and $\textbf{u}$ are the density and velocity of the gas at a certain location. Integrating over a spherical volume of radius $R$,
\begin{equation}
\frac{dM}{dt} = -\oiint \rho \textbf{u}\cdot \,\mathrm{d}\textbf{S},
\label{eq:continuity}
\end{equation}
where $M$ is the total gas mass inside the volume and the last integral is evaluated on the enclosing spherical surface, with ${\mathrm{d}S=R\sin\theta \,\mathrm{d}\theta \,\mathrm{d}\varphi}$. The flux can be measured in two different ways: (i) as $(1/4\pi R^2)\delta M/\delta t$, \ie\ using the difference, between two snapshots, of the total gas mass inside the sphere (thus averaging over $\delta t$); or (ii) estimating the integral on the right-hand side of Eq.~\ref{eq:continuity} in a shell of small but finite thickness $\delta R$ (thus averaging over $\delta R$). The first method presents practical problems because, at the centre of the volume, a fraction of gas will form stars. Moreover, we are interested in the gas \textit{inflow} along the warp, as opposed to total mass variation, which includes feedback-driven outflows. Finally, we are interested in the frequencies associated with the variation of the gas flux; thus we opt to use single snapshots at each time, \ie\ the second method.

We compute the right-hand side of Eq.~\ref{eq:continuity} via a Monte Carlo integration. For this, we introduce the function $f$, which represents the underlying probability distribution from which particles in the shell are sampled, such that $\int f \,\mathrm{d}S = 1$. The right-hand side of Eq.~\ref{eq:continuity} is estimated as
\begin{equation}
I\equiv -\oiint \frac{\rho \textbf{u}}{f}f\cdot \,\mathrm{d}\textbf{S} \approx -\frac{1}{N} \sum_i \frac{\rho_i u_{ri}}{f_i},
\label{eq:I_int}
\end{equation}
where we sum over gas particles in the shell ($N$ particles), $\rho_i$ is the density around particle $i$, $u_{ri}$ is its radial velocity component and we select cool gas ($T_g\leq50,000$ K) particles with $u_{ri} <0$. The sampling function $f$ is obtained marginalising over the number density profile, $n(R,\theta,\varphi)$, within the shell,
\begin{equation}
f(\theta,\varphi|R) = \frac{1}{N}\int_{R-\delta R/2}^{R+\delta R/2}n(R',\theta,\varphi) \, \mathrm{d}R' \approx \frac{1}{N}n(R,\theta,\varphi)\delta R.
\end{equation}
Substituting in Eq.~\ref{eq:I_int}, we finally estimate the flux as
\begin{equation}
\frac{I}{4\pi R^2}\approx -\frac{1}{4\pi R^2 \delta R} \sum_i m_i u_{ri},
\end{equation}
where $m_i$ is the mass of particle $i$ and we approximated the density around the particle by the smooth density profile evaluated at the center of the shell, \ie\ $\rho_i\approx \rho(R,\theta,\varphi)$.
	
The top panel of Fig.~\ref{fig:bubble} shows the evolution of the inward mass flux of cool gas ($T_g<50,000$ K) through a spherical shell with thickness $\delta R=0.2 \kpc$ and radius $R=15 \kpc$. The flux of cool gas varies substantially, with long term inflow modulated by rapid variations. Similar to the analysis in Section \ref{sec:spectral}, we apply a discrete Fourier transform to the evolution of the mass flux over $1 \Gyr$ baselines to derive the characteristic timescales of the variations. The bottom panel of Fig.~\ref{fig:bubble} shows the resulting frequencies of the mass flux. Most of the frequencies cluster between $0$ and $20~\kmsk$, and reaching to $40~\kmsk$. These results show that the disc is continuously perturbed by the irregularly accreting gas with a maximum amplitude of $5.8 \times 10^{6} \Msun \kpc^{-2} \Gyr^{-1}$, and typical amplitudes of $\sim 10^{6} \Msun \kpc^{-2} \Gyr^{-1}$, which is comparable to recent estimates of the gas inflow in the Milky Way \citep{gas_inflow_est, Werk+2019}.
These frequencies substantially overlap the frequencies of the bending waves, indicating a favourable spectrum of perturbations for exciting the bending waves. We propose, therefore, that the irregular inflow of gas from the warp onto the disc is the source of the vertical perturbations which excite the bending waves in the warped model. 
	
\begin{figure}
		\includegraphics[width=1.\linewidth]{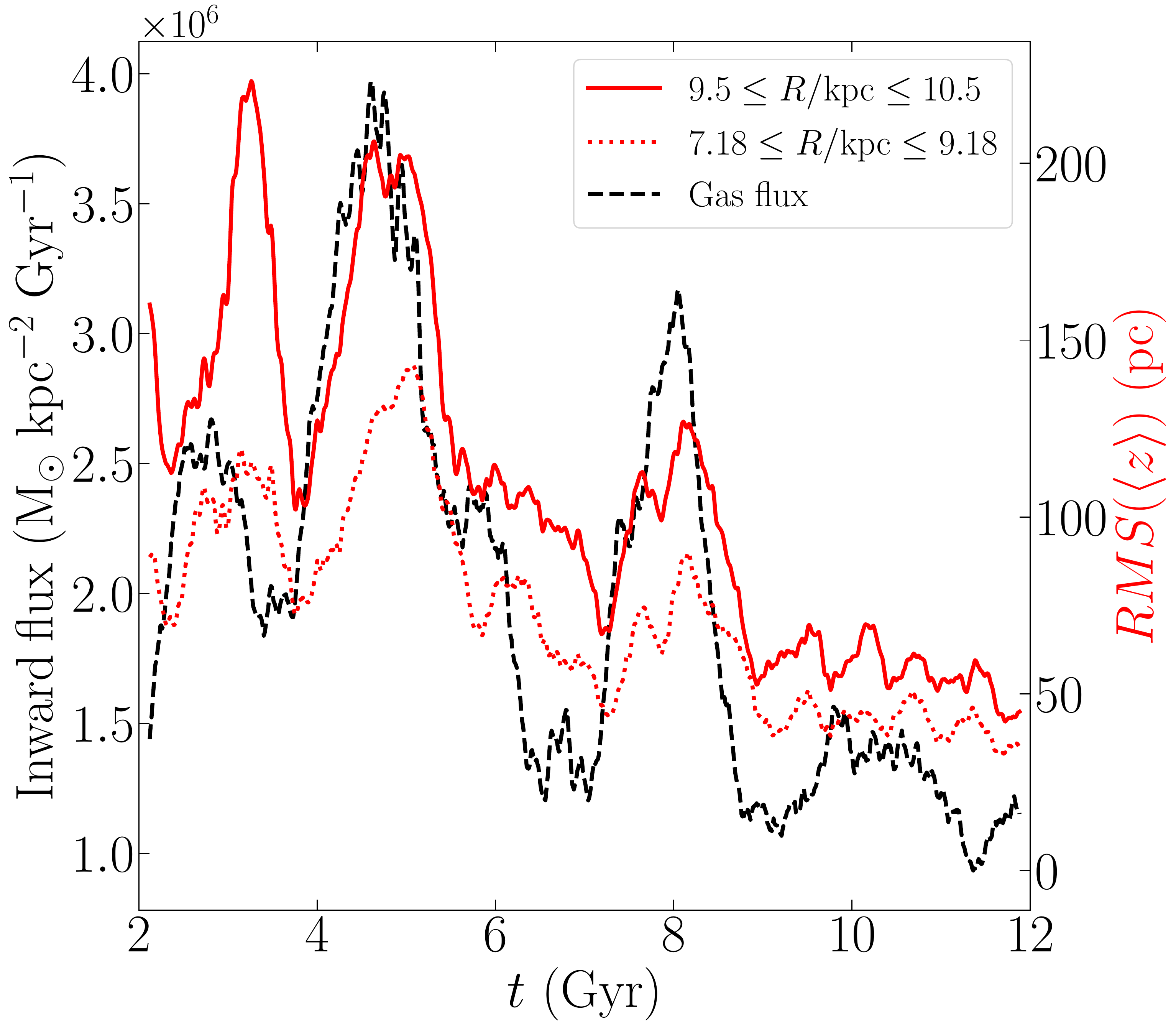}  
		\caption{Evolution of the gas flux presented in Fig.~\ref{fig:bubble} (black, left $y$-axis) and of the root-mean-square of the mean vertical displacement, $RMS( \avg{z})$ (red, right $y$-axis), at $7.18\leq R/\kpc\leq9.18$ (Solar annulus, dotted) and $9.5\leq R/\kpc\leq 10.5$ (outer disc, solid). $RMS( \avg{z})$ is calculated using the azimuthally binned \avg{z} values in  Figs.~\ref{fig:z_vzed_evolution_ws} (left column) and \ref{fig:zvzed_slope} (top panel) for the outer disc and Solar annulus, respectively. The gas flux and $RMS( \avg{z})$ are averaged over $0.25 \Myr$ intervals. A cross-correlation analysis of these different series reveals time lags, compared to the gas flux, of $200 \Myr$ and $250 \Myr$ at the outer and inner annuli, respectively (see main text for details).
		}
		\label{fig:flux_zed}
	\end{figure}
	
To further demonstrate that cold gas accretion has direct impact on the vertical structure of the galactic disc, we analyse the evolution of the gas flux relative to the total vertical power at different annuli.
We estimate the total vertical power by calculating \avg{z} in sectoral non-overlapping bins with $\Delta\phi_w=12\degrees$ at each annulus and then taking the root-mean-square ($RMS$) across the azimuthal bins.

Fig.~\ref{fig:flux_zed} shows the evolution of the flux (black) and $RMS( \avg{z})$ (red) at the Solar annulus (dotted) and the outskirts of the disc (solid). Measuring the cross-correlation between the flux and $RMS( \avg{z})$ shows that there is a lag of $\sim 200 \Myr$ and $\sim 250 \Myr$ at the outer disc and Solar annulus, respectively. The overall lag is expected as the gas flux is measured at $R=15\kpc$ so it takes time to reach and impact the disc. The $\sim 50 \Myr$ lag between peak in the outer disc and that at the Solar annulus represents the time required for the excited waves to propagate from the outskirts to the Solar annulus. Considering the distance $\Delta R\sim1.8\kpc$, this signal propagates with velocity $\sim - 36 \pc \Myr^{-1}$ (the minus sign indicating inward propagation).

We can now link this velocity with the expected group velocity of bending waves, $d\omega/dk$. For a simple estimate, we use the WKB dispersion relation for an $m=1$ bending wave \citep{Toomre83}
\begin{equation}
    \label{eq:HT_DR}
    \left[\omega - \Omega(R)\right]^2 - 2\pi G\Sigma(R)|k| - \nu_h^2 = 0,
\end{equation}
where $k$ is the wave number and $\Sigma(R)$ is the surface density. The expected group velocity is then
\begin{equation}
    \label{eq:group_vel}
    \left.\frac{d\omega}{dk}\right|_R = \frac{\rm{sgn}(k)\pi G\Sigma(R)} {\Omega_\mathrm{p}-\Omega(R)},
\end{equation}
according to which a negative group velocity can be associated with a leading ($k<0$) prograde wave, or to a trailing ($k>0$) retrograde wave. The spectral analysis showed an overall larger power in the retrograde wave, and Fig.~\ref{fig:z_ws_uw} suggests a trailing shape. Substituting the value $\Omega_p=-12.8~\kms\kpc^{-1}$ identified in Fig.~\ref{fig:670h_bubble} at late times, and $\Sigma = 80\Msun\pc^{-2}$ and $\Omega=18.6 \kms\kpc^{-1}$ (values at $R\sim 9.1 \kpc$), we obtain a radial group velocity of $\sim -35 \pc \Myr^{-1}$, in striking agreement with the value obtained from the measured time lag. This agreement is somehow surprising given all the approximations and simplifying assumptions involved, so it should be considered with caution.

\subsection{Vertical kinematics in the SN}
\label{subsec:sch}
	
	\begin{figure*}
		\includegraphics[width=.8\linewidth]{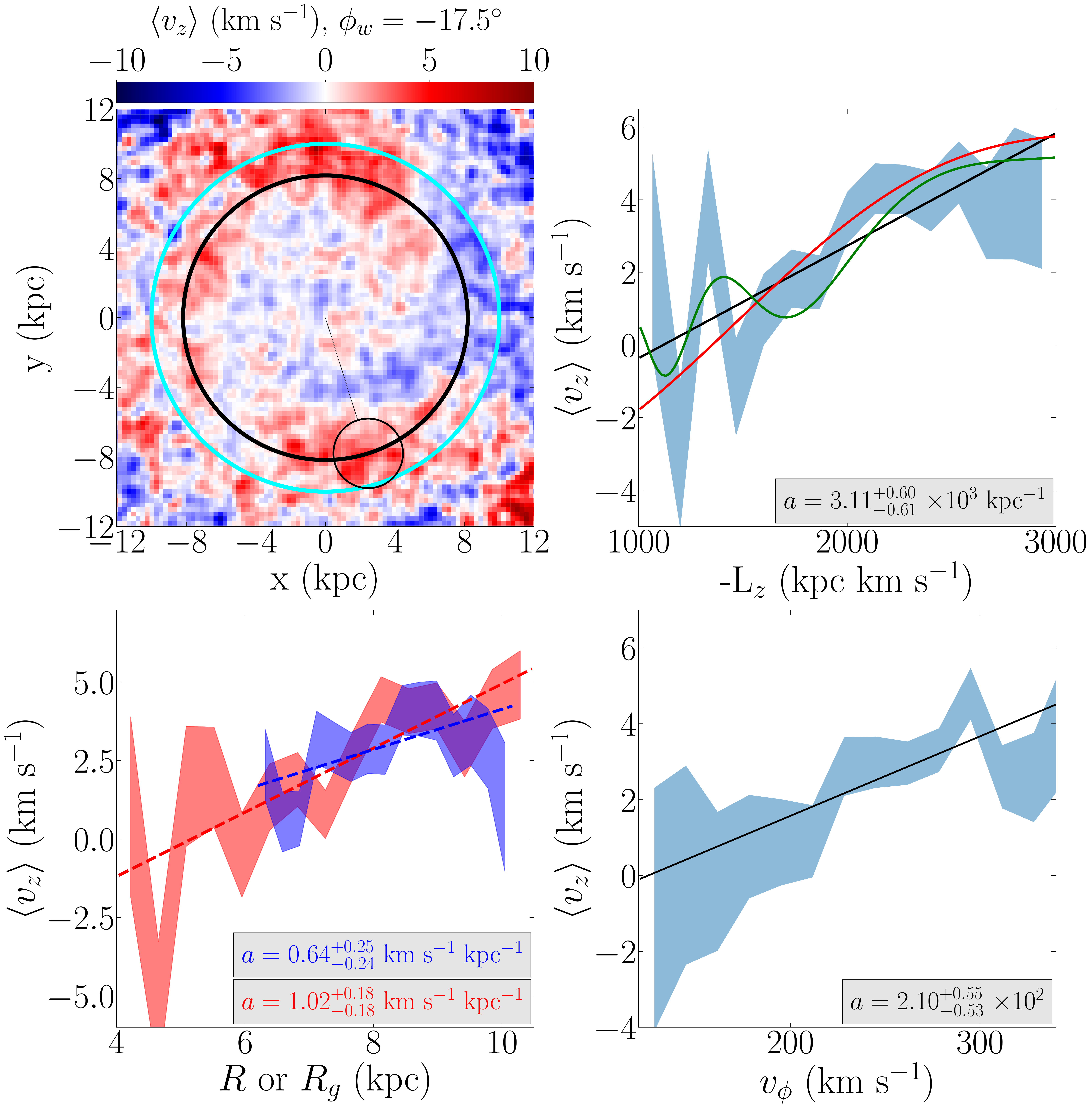}
		\caption{
			Top left: Distribution of \avg{v_z} in the stellar disc of the warped model at $11.4\Gyr$. The smaller black circle indicates a region of radius $2\kpc$ to simulate a SN (azimuth indicated above the colour bar). The larger solid black and cyan circles represent the Solar annulus, $R=8.18\kpc$, and $R=10\kpc$, respectively. A Gaussian filter has been applied to the colour distribution with a standard deviation set to $\sigma=1$ pixel = $260\times 260 \pc$.
			Top right: binned distribution of $\avg{v_z}$ as a function of the angular momentum, $L_z$, in the SN sample of the warped model. The shaded regions show the standard deviation of \avg{v_z} in each bin. There are three model fits present: linear (Eqn.~\ref{eq:lin}, black), sinusoidal (Eqn.~\ref{eq:sin}, red), and wrapping (Eqn.~\ref{eq:wrp}, green). The slope of the fitted linear model is shown in the panel's bottom right corner, while the rest of the fit parameters for the linear and other models are presented in Table \ref{tab:my-table}. Bottom left: binned distributions of $\avg{v_z}$ as functions of cylindrical radius, $R$ (blue) and guiding radius, $R_g$, (red) in the same SN sample. Each distribution has a fitted linear model (dashed lines). Bottom right: binned distribution of $\avg{v_z}$ as a function of $v_\phi$ in the same SN sample, as well as the fitted linear model. The respective slope values are shown in the bottom right corner of each panel. The choice of time step is not arbitrary; at $11.4 \Gyr$ is the last moment the SN sample has a slope of $a >3\times10^3 \kpc^{-1}$ (see Fig.~\ref{fig:slopes_uw_ws} for details).
			}
		\label{fig:vz_distr_ws}
	\end{figure*}
	
	In Fig.~\ref{fig:z_ws_uw} the vertical bends are accompanied by non-zero \avg{v_z}. \citetalias{wrinkle} and \citetalias{huang} observed an increase in \avg{v_z} with angular momentum $|L_z|$, which they speculated was due to either an extension of the warp, or to a bending wave. We test whether such signals arise in our models.
	
	Fig.~\ref{fig:vz_distr_ws} examines a simulated SN sample in the warped model at $11.4 \Gyr$, with plots similar to those of \citetalias{wrinkle} and \citetalias{huang}. Three panels plot $\avg{v_z}$ versus $L_z$ (top right), versus azimuthal velocity, $v_{\phi}$ (bottom right), and versus cylindrical and guiding radii, $R$ and $R_g$ \citep[bottom left, the latter computed using {\sc agama},][]{agama}. With the improved mapping of the Milky Way's warp, the Sun's position relative to it is now clearer: the Sun is $\sim 17.5\pm1\degrees$ behind the ascending node of the warp \citep{chen}. Our sample is contained within a sphere of radius $2\kpc$ at $R = 8.18\kpc$ and azimuth $\phi_w=-17.5\degrees$, where $\phi_w$ is the azimuthal angle along the direction of rotation measured from ascending node of the warp's LON. The location of our sample is indicated in the top left panel of Fig.~\ref{fig:vz_distr_ws}, on top of a face-on map of $\avg{v_z}$. Although all of the binned \avg{v_z}\ variations have relatively larger errors (despite our bins being large compared with \citetalias{wrinkle} and \citetalias{huang}), we observe a general increase of \avg{v_z} with $-L_z$ along with underlying wiggles, as in the Milky Way.
	Following \citetalias{wrinkle}, we fit a variety of functions to the $\avg{v_z}$ versus $L_z$ distribution:
	\begin{equation}
	\label{eq:lin}
	\avg{v_z} = b + aL^{\prime}_{z},
	\end{equation}
	\begin{equation}
	\label{eq:sin}
	\avg{v_z} = b + aL^{\prime}_{z}+A\sin(2\upi L^{\prime}_{z}/c+d),
	\end{equation}
	and
	\begin{equation}
	\label{eq:wrp}
	\avg{v_z} = b + aL^{\prime}_{z}+A\sin(2\upi c/ L_{z}+d),
	\end{equation} 
	where $L_z^{\prime}=L_z-1600 \kmsk$ in the Milky Way, and $a$, $b$, $c$, $d$, $A$ are fitting parameters. For the warped model we set $L_z^{\prime}=L_z-2000 \kmsk$ based on the mean value of $L_z$ at $R=8.18\kpc$, but note that in the fit of Eqn.~\ref{eq:lin} the slope is independent of this pivot point. Assuming $v_z$ is normal-distributed and using flat priors for all parameters, we sample the posterior distribution function with the {\sc emcee} package \citep{Foreman-Mackey+2019}. The best fit parameters for Eqns.~\ref{eq:lin}~-~\ref{eq:wrp} are listed in Table~\ref{tab:my-table} (with uncertainties estimated as the interval containing $68\%$ of samples around the median), and we see that our linear fit is of the same scale as the one measured in the Milky Way.
	The fits to Eqns.~\ref{eq:sin}~-~\ref{eq:wrp} present larger uncertainties and deviations from the values of \citetalias{wrinkle} and \citetalias{huang}, thus we mainly focus on the fits of the simple linear function. We also perform linear fits on the other $\avg{v_z}$ distributions (versus $v_\phi$, $R$, and $R_g$), the slopes of which are presented in the bottom right corner of the respective panels. These additional fits are also of the same order as those measured in the SN \citepalias{wrinkle,huang}.
	
	\begin{table*}
		\centering
		\caption{Best fit parameters for the fitting models of Eqns. \ref{eq:lin}~-~\ref{eq:wrp} applied to the sample in Fig.~\ref{fig:vz_distr_ws}. The slope of the linear fit has been measured in the SN by \protect{\citetalias{wrinkle}} ($3.05\pm0.25\times10^3\kpc^{-1}$) and \protect{\citetalias{huang}} ($3.11\pm0.70\times10^3\kpc^{-1}$) which is within the range we find.}
		\label{tab:my-table}
		\begin{tabular}{lcccccllll}
			\hline
			\multicolumn{1}{c}{Fit} & \begin{tabular}[c]{@{}c@{}}a\\ ($\times 10^{3}$ kpc$^{-1}$)\end{tabular} &                      & \begin{tabular}[c]{@{}c@{}}b\\ (km s$^{-1}$)\end{tabular} &                      & \begin{tabular}[c]{@{}c@{}}c\\ (kpc km s$^{-1}$)\end{tabular} & \multicolumn{1}{c}{} & \multicolumn{1}{c}{\begin{tabular}[c]{@{}c@{}}d\\ $\;$\end{tabular}} & \multicolumn{1}{c}{} & \multicolumn{1}{c}{\begin{tabular}[c]{@{}c@{}}A\\ (km s$^{-1}$)\end{tabular}} \\ \hline
			linear  (Eq. \ref{eq:lin})                 & $3.11 ^{+0.59}_{-0.58}$                                                      &                      & $2.72 ^{+0.24}_{-0.24}$                                              &                      & -                                                               & \multicolumn{1}{c}{} & \multicolumn{1}{c}{-}                                                & \multicolumn{1}{c}{} & \multicolumn{1}{c}{-}                                                          \\
			& \multicolumn{1}{l}{}                                              & \multicolumn{1}{l}{} & \multicolumn{1}{l}{}                                       & \multicolumn{1}{l}{} & \multicolumn{1}{l}{}                                            &                      &                                                                      &                      &                                                                                \\
			sinusoidal (Eq. \ref{eq:sin})                 & $2.96 ^{+0.89}_{-1.4}$                                                      &                      & $2.41 ^{+0.51}_{-5.6}$                                               &                      & $3443.59 ^{+33459.11}_{-3073.73}$                                              &                      & $0.73 ^{+1.10}_{-1.72}$                                                        &                      & $1.24 ^{+6.89}_{-0.80}$                                                                  \\
			& \multicolumn{1}{l}{}                                              & \multicolumn{1}{l}{} & \multicolumn{1}{l}{}                                       & \multicolumn{1}{l}{} & \multicolumn{1}{l}{}                                            &                      &                                                                      &                     	 &                                                                                \\
			wrapping (Eq. \ref{eq:wrp})                 & $2.72 ^{+0.66}_{-0.66}$                                                      &                      & $2.56 ^{+0.25}_{-0.25}$                                               &                      & $3601.17 ^{+151.99}_{-323.90}$                                              &                      & $-2.59 ^{+1.05}_{-0.42}$                                                        &                      & $1.10 ^{+0.39}_{-0.42}$                                                           \\ \hline
		\end{tabular}
	\end{table*}

	\begin{figure}
		\includegraphics[width=1.\linewidth]{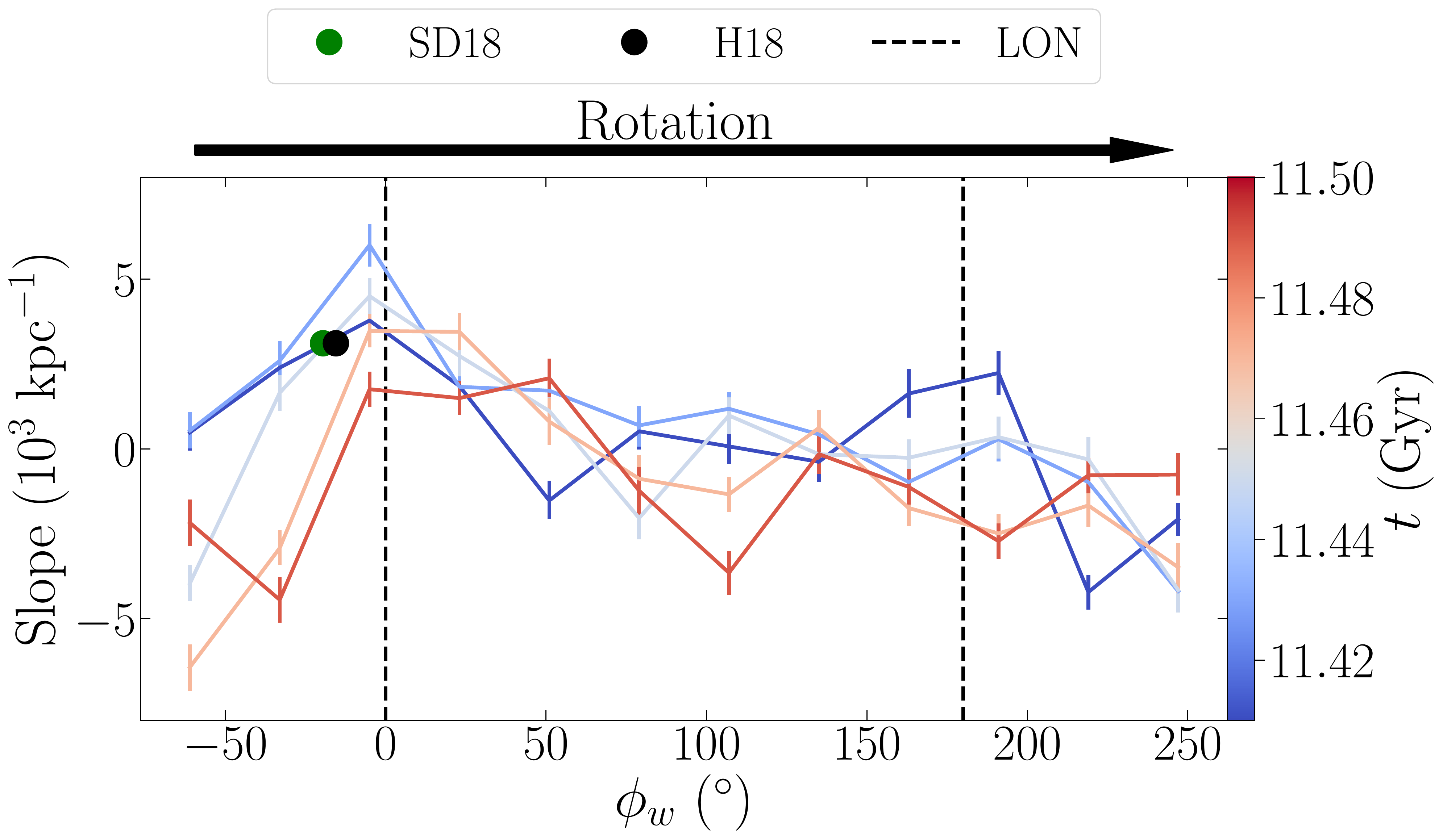}
		\caption{The slope of the $\avg{v_z}$-$L_z$ relation as a function of azimuth for samples at $R = 8.18\kpc$ in the warped model. The value $\phi_w = 0$ is defined as the azimuth on the ascending node of the warp's LON (dashed line). Therefore the descending node, similarly indicated by a vertical dashed line, is at $180\degrees$. The sense of rotation is indicated above the figure. In the Milky Way, the Sun is located $17.5\degrees$ before the ascending node \protect\citep{chen}, \ie\ at $\phi_w = -17.5\degrees$.  The black and green dots represent the slope as measured by \protect\citetalias{wrinkle} and \protect\citetalias{huang}, respectively (horizontally offset by $\pm 2\degrees$ for clarity).  The panel shows 5 snapshots separated by $20\Myr$. Waves are seen propagating in the direction of rotation, \eg\ starting at $50\degrees$ (dark blue) and reaching $110\degrees$ (dark red).}
		\label{fig:slope_azimuth_ws}
	\end{figure}	
	
	\citetalias{wrinkle} argued that one possible interpretation of the non-vanishing slope of the \avg{v_z}-$L_z$ relation is that the stellar disc is warped at the Solar cylinder. The slope of the relation in this scenario would vary smoothly with azimuth as $\cos(\phi_w + \phi_c)$ (where $\phi_c$ is some constant), which we can check in our model. In Fig.~\ref{fig:slope_azimuth_ws} we plot the slope, $a$, of the linear fit of Eqn.~\ref{eq:lin} as a function of $\phi_w$, the azimuthal angle at which the sample is selected. This relation is plotted for a number of snapshots, with a time interval $\delta t = 20 \Myr$ to show the short-term changes in the slope. For this measurement we use 12 samples that consist of $2 \kpc$ spheres. The spheres are equally spaced in azimuth to avoid overlapping the samples. The slope varies in the range $[-5,5]\times10^{3}\kpc^{-1}$; \citetalias{wrinkle} and \citetalias{huang} measure a slope of $\sim 2.64-3.21 \times 10^3\kpc^{-1}$, which is within the range we find. The results of these snapshots happen to be instances when the slope at the Solar azimuth is very similar to that observed in the Milky Way. Note that $a$ varies in a wave-like manner as the peaks and valleys shift with time. As the warp is fixed at each snapshot (see Section~\ref{ssec:preproc}) the positive slope in the \avg{v_z}-$L_z$ relation is not produced by the warp itself, but by a propagating bending wave, which suggests that the same may be happening in the MW. The phase of the wave moves in the direction of increasing $\phi_w$, \ie\ in the sense of rotation.
	
	\begin{figure*}
		\includegraphics[width=.8\linewidth]{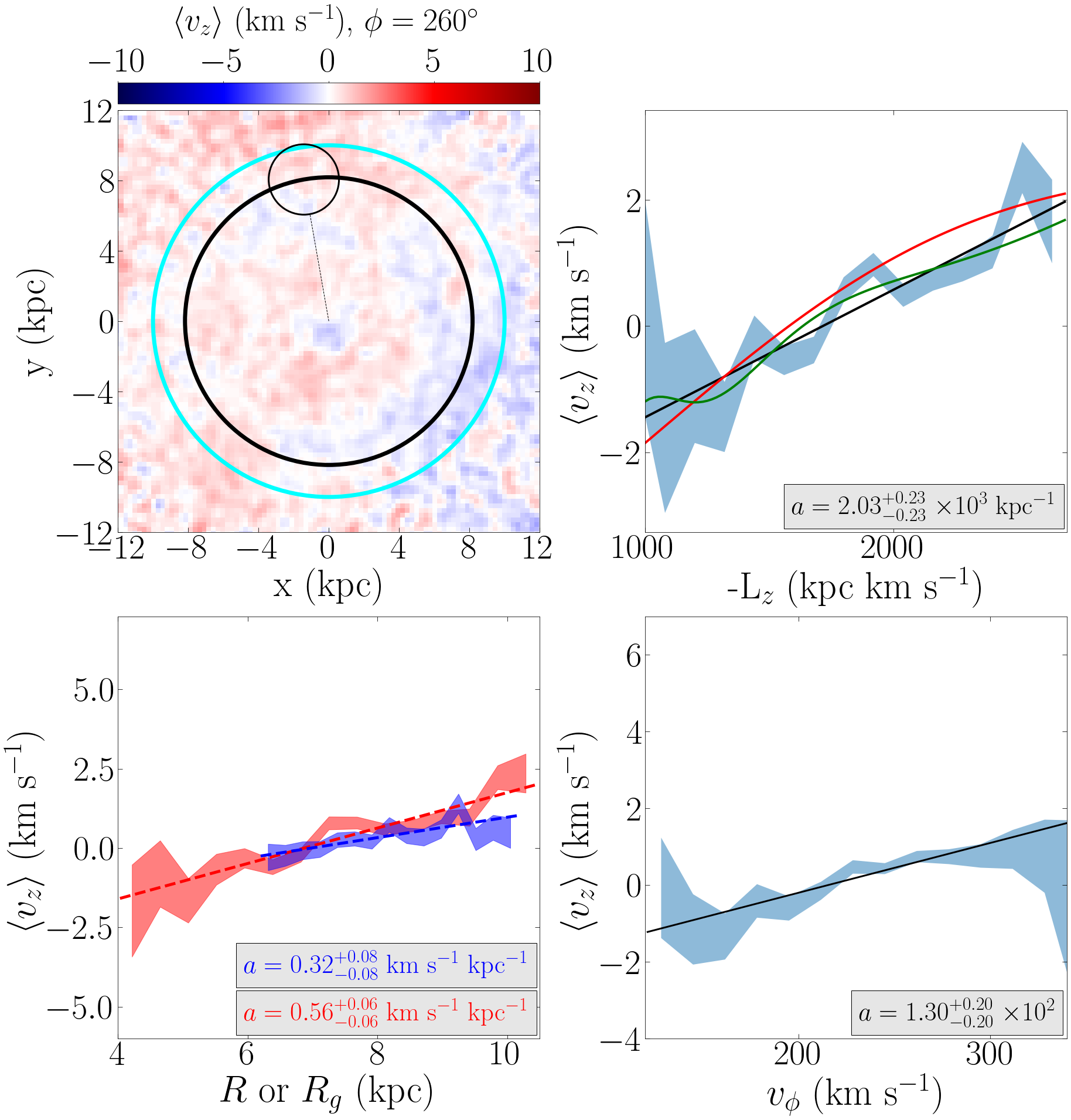}
		\caption{Similar to Fig.~\ref{fig:vz_distr_ws} but for the unwarped model at $11.8\Gyr$. Top left: Distribution of \avg{v_z} of the stellar disc. The small black circle indicates a region of radius $2\kpc$ to simulate a SN (azimuth indicated above the colour bar). The solid black and cyan lines represent the Solar annulus, $R=8.18\kpc$, and $R=10\kpc$, respectively. A Gaussian filter has been applied to the colour distribution with a standard deviation set to $\sigma=1$ pixel = $260\times 260 \pc$. Top right: binned distribution of $\avg{v_z}$ as a function of the angular momentum, $L_z$, in the SN sample of the unwarped model. The shaded region shows the standard deviation of \avg{v_z} in each bin. There are  three model fits present: linear (black line), sinusoidal (red line), and wrapping (green line). Bottom left: binned distributions of $\avg{v_z}$ as functions of radii, $R$, (blue lines) and guiding radii, $R_g$, (red lines) in the same SN sample. Each distribution has a fitted linear model (dashed lines). Bottom right: binned distribution of $\avg{v_z}$ as a function of $v_\phi$ (blue lines) in the same SN sample. The distribution has a fitted linear model (black line). The slope of the fitted linear models are shown in the bottom right corners of the respective panels.
		}
		\label{fig:vz_distr_uw}
	\end{figure*}

	\begin{figure}
		\includegraphics[width=1.\linewidth]{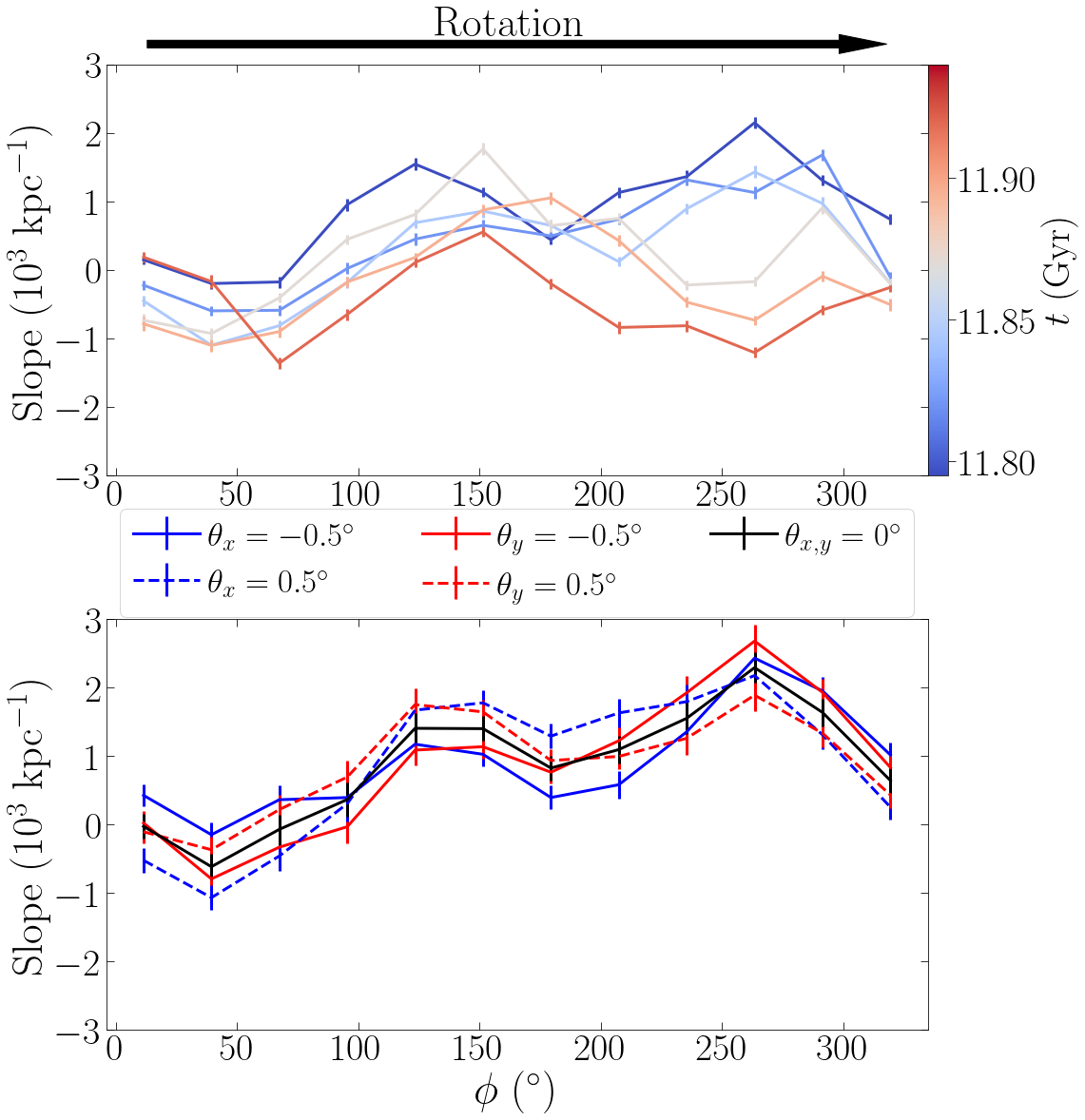}
		\caption{Slope of the $\avg{v_z}$-$L_z$ relation with azimuth for SN samples in the unwarped model. Top panel: 5 times separated by $20\Myr$ (colours). Bottom panel: slope variation, for the unwarped model at $11.8\Gyr$, when the largest slope is observed, with different artificial tilts about the $x$-axis (red lines), about the $y$-axis (blue lines), and without any artificial tilt (black line). Note that the range of the $y$-axis for both panels is almost a third of that in Fig.~\ref{fig:slope_azimuth_ws}. This demonstrates that if the midplane were defined inaccurately, the resulting small tilts would not produce the large slopes measured in the Milky Way or in the warped model.}
		\label{fig:slope_azimuth_uw}
	\end{figure}

	Fig.~\ref{fig:vz_distr_uw} plots the variation of \avg{v_z} with $L_z$ (top right), $v_\phi$ (bottom right), and $R$ and $R_g$ (bottom left) in the unwarped model. As the simulation is unwarped the simulated SN is arbitrary, so we perform our analysis at $R=8.18\kpc$ in 30 different azimuths and present the sample with the largest recent slope. The relation is significantly shallower than in the warped simulation. Fig.~\ref{fig:slope_azimuth_uw} shows the variation of the slope with azimuthal angle at different times (top panel), similar to Fig.~\ref{fig:slope_azimuth_ws}. The variation of the slope with azimuth is less pronounced when compared to the warped model and barely reaches the Milky Way values throughout the $2\Gyr$ interval, with $|a| \la 2 \times 10^3 \kpc^{-1}$. The bottom panel shows the effect of small artificial tilts of the disc about the $x$ (red) and $y$ (blue) axes. These small $(0.5\degrees)$ tilts barely change the slope, indicating that the large slope observed in the Milky Way is not due to a mis-identified disc mid-plane.
	
	\begin{figure}
		\includegraphics[width=1.\linewidth]{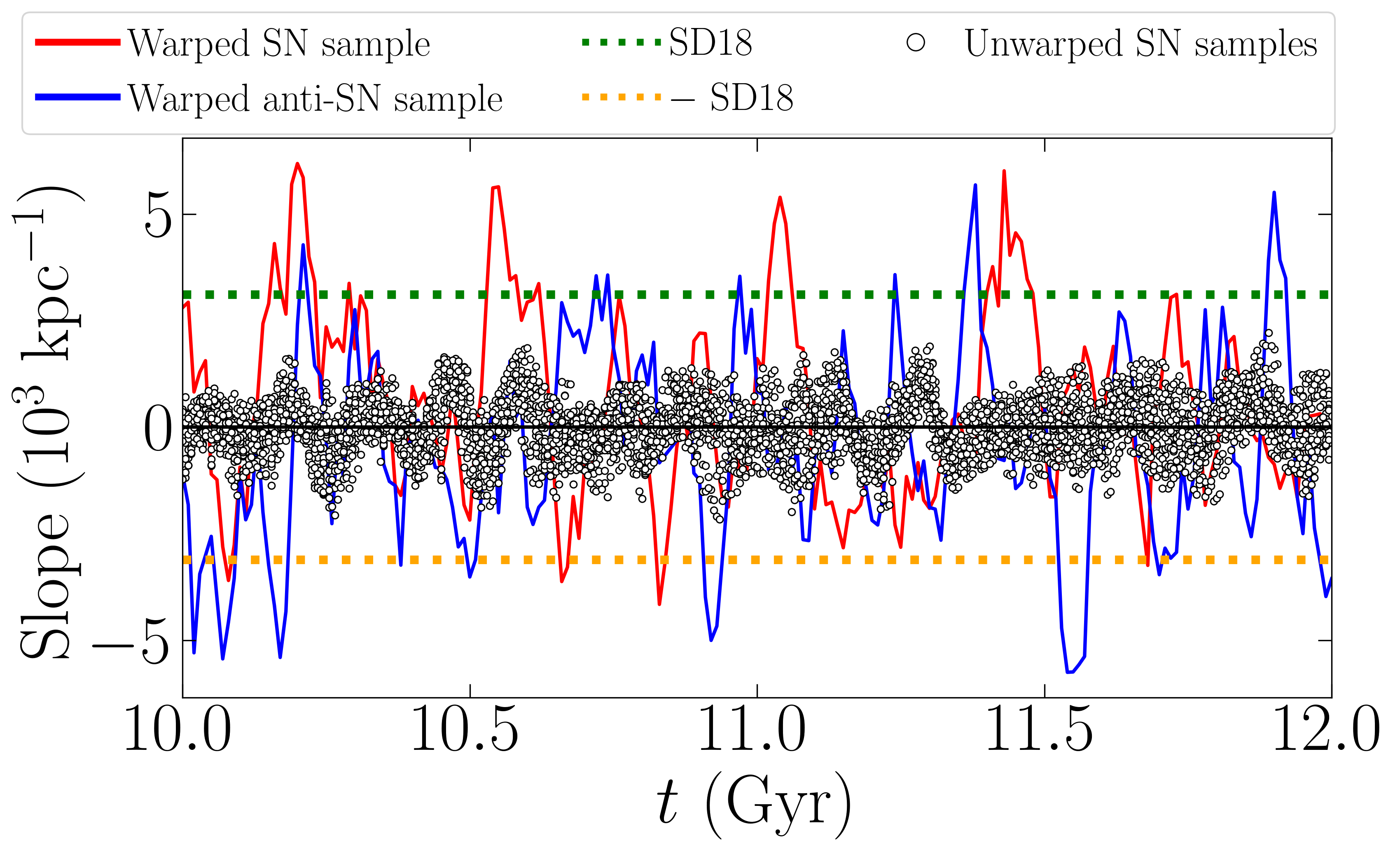}
		\caption{Evolution of the slope of the $\avg{v_z}$-$L_z$ relation for all SN samples in the unwarped model (open circles) and SN samples in the warped model at $\phi_w=-17.5\degrees$ (red) and $\phi_w=162.5\degrees$ (blue). The samples are spheres centred on $R=8.18\kpc$ and with $r=2 \kpc$. The green dotted line shows the SN slope value \protect\citepalias{wrinkle}, while the orange dotted line is the negative of that value. In the unwarped model the mean and overall slope values do not generally exceed $2\times10^3\kpc^{-1}$ in the span of $2 \Gyr$ and at any SN sample. In the warped model the slope regularly matches, or exceeds, the MW value.}
		\label{fig:slopes_uw_ws}
	\end{figure}	
	
Since the unwarped simulation lacks a line of nodes to simulate a SN sample, as in the analysis of Fig.~\ref{fig:slope_azimuth_ws}, we measure the slope in 12 azimuthally equally spaced $2 \kpc$ spheres. Fig.~\ref{fig:slopes_uw_ws} plots the slope values for these 12 samples (white points) over a $2\Gyr$ interval starting from $t=10\Gyr$ with $\delta t=10\Myr$. The slope values oscillate about $a=0 \kpc$ without reaching the SN values (green dotted line). In contrast, the evolution of the slope in the warped model's SN sample (red solid line) shows strong oscillations about $a\sim0.6 \times 10^3\kpc^{-1}$ with more than half of the values being positive. There are multiple time intervals ($\sim15\%$ of time steps) where the slope reaches and surpasses the \protect\citetalias{wrinkle} and \protect\citetalias{huang} values. 
	
The Sun is located behind the ascending node of the warp \citep{chen}, which could have an impact on the $L_z$ vs $v_z$ relation. In order to explore how the Sun's location relative to the line of nodes affects the measured slope, we measure the slope and its evolution in an ``anti" SN (anti-SN) sample. The sample is located behind the descending node of the warp, \ie\ $\phi_w=162.5\degrees$ (blue solid line in Fig.~\ref{fig:slopes_uw_ws}). We observe that the slope at the anti-SN location oscillates about $a\sim-0.5 \times 10^3\kpc^{-1}$. More than half of the slope values are now negative and the \citetalias{wrinkle} and \citetalias{huang} values are reached (or exceeded) in only a third of the time as the SN sample ($\leq6\%$ of time steps). However, when considering the negative of the slope in \citetalias{wrinkle} and \citetalias{huang} (orange dotted line), the anti-SN sample reaches that value at the same rate as the SN sample reaches the real value.

We conclude that the bending waves produced by misaligned gas accretion along the warp in the simulation are able to produce similar trends as found by \citetalias{wrinkle} and \citetalias{huang} in the MW. The large positive values of the slope found by \citetalias{wrinkle} and \citetalias{huang} are not unusual given the Sun's position relative to the line-of-nodes of the warp.

\subsection{Propagation of the bending waves}
\label{ssec:bends}
\begin{figure*}
		\includegraphics[width=1\hsize]{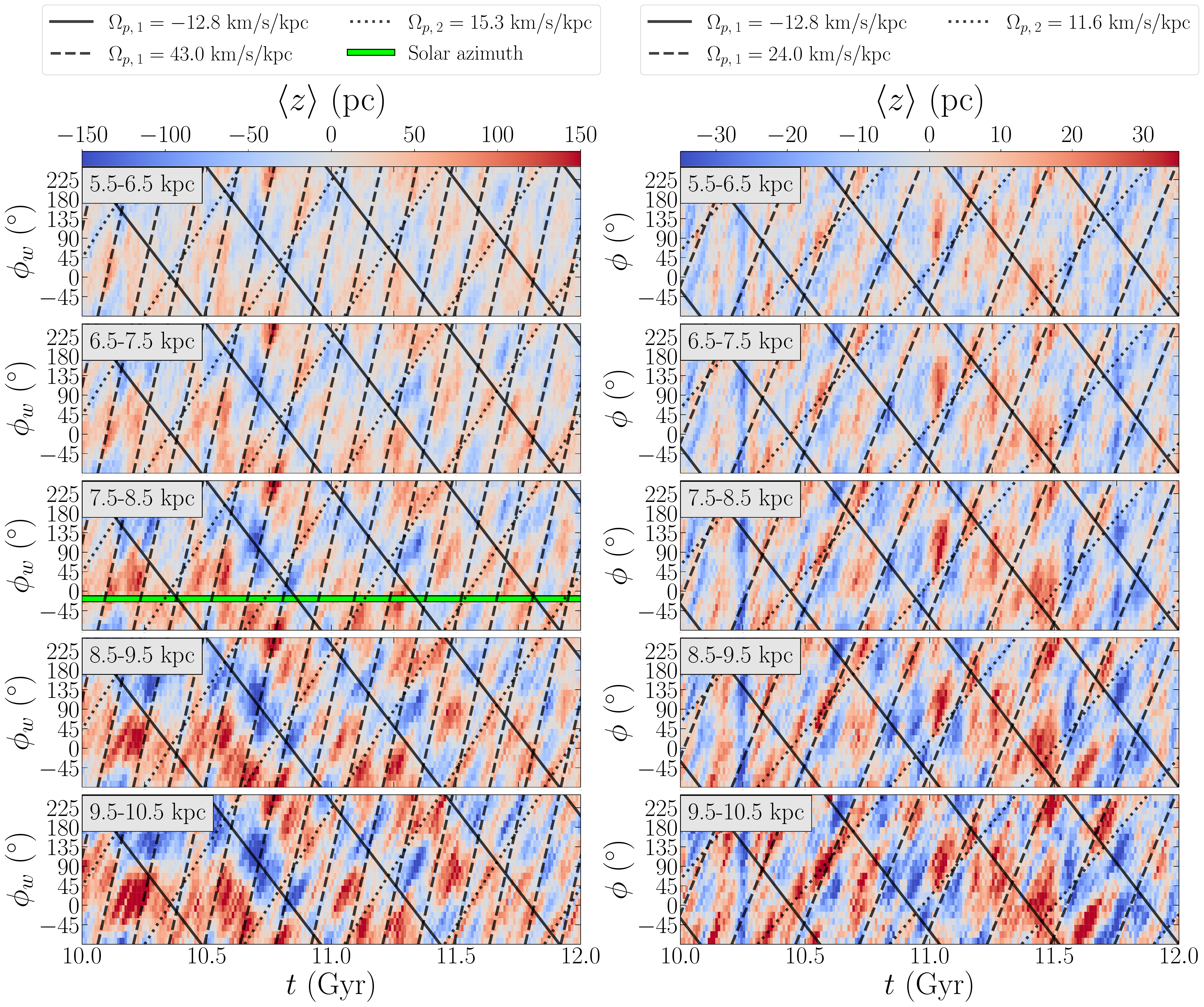}
		\caption{Evolution of the stellar mean height above the mid-plane, \avg{z}, in the warped (left) and unwarped (right) simulations.  The horizontal solid green line represents the Solar azimuth relative to the ascending node of the warp's LON \protect\citep{chen}. The diagonal black lines correspond to the most prominent retrograde $m=1$ (solid) and prograde $m=1$ (dashed) and $m=2$ (dotted) bending pattern speeds present in the $10-12\Gyr$ interval (see Figs.~\ref{fig:739HF_hc_bubble} and ~\ref{fig:670h_bubble}).}
		\label{fig:z_vzed_evolution_ws}
\end{figure*}

In Sections~ \ref{sec:spectral} and \ref{subsec:sch} we established the presence of bending waves, their pattern speeds throughout the disc, and their consequences on vertical kinematics in a simulated SN. We now explore the time evolution of \avg{z} distribution of stars in cylindrical sectors. We use sectoral bins which are non-overlapping with $\Delta\phi_w = 12\degrees$ in each ring. This analysis is a counterpart of the frequency analysis but in real space, which aids in understanding the observational consequences of the waves found in Sec.~\ref{sec:spectral}. 

\subsubsection{Warped simulation}

For the warped simulation, the left column of Fig.~\ref{fig:z_vzed_evolution_ws} shows the evolution of \avg{z}\ in $1\kpc$ wide rings from $5.5\kpc$ to $10.5\kpc$ (rows), starting at $t=10\Gyr$ with time steps $\delta t=10\Myr$. The horizontal green line shows the location of the SN in the Milky Way \citep{chen}. The diagonal black lines are the most prominent frequencies of the bending waves taken from Figs.~\ref{fig:670h_psp} (bottom right) and~\ref{fig:670h_bubble} (right) for this time interval. The values of the frequencies are indicated in the legend at the top of Fig.~\ref{fig:z_vzed_evolution_ws}.

The distributions of \avg{z} at any time are dominated by an $m=1$ angular dependence, \ie\ at each annulus and time interval \avg{z}\ has a single peak (red) and trough (blue).\footnote{We remind the reader of the equivalence of the angles $360\degrees$ and $0\degrees$, \ie\ that the top end of each panel wraps back to the bottom.}
Over time the bend propagates in a retrograde direction, \ie\ in the direction of decreasing $\phi_w$. The pattern speed of the retrograde bending wave (solid, black line), identified in Fig.~\ref{fig:670h_bubble}, matches the slope of the $m=1$ signal in the \avg{z} distribution at all radii. This coincidence of slopes is in agreement with Fig.~\ref{fig:670h_psp}, which shows that the $\Omega_{p}\approx-13\kms\kpc^{-1}$ bending wave is spread across those radii.
	
Superposed on the overall $m=1$ bending wave we can also see individual bending wave packets (which we loosely refer to as ``ripples" to distinguish their particular behaviour) which propagate to {\it increasing} $\phi_w$, \ie\ in the direction of rotation. These ripples are probably the result of constructive interference between the more prominent retrograde $m=1$ bending waves and the less powerful prograde $m=1$ and $m=2$ bending waves. In Fig.~\ref{fig:670h_psp}, we identify the prominent prograde $m=1$ ($\Omega_{p}\approx43\kms\kpc^{-1}$) and $m=2$ ($\Omega_{p}\approx15\kms\kpc^{-1}$) bending wave pattern speeds and overlay them on top of the \avg{z} maps at each radius. The prograde pattern speeds appear to match the slopes of the ripples at each radial interval. Figs.~\ref{fig:670h_psp} and \ref{fig:670h_bubble} demonstrate that there are more frequencies present in the warped simulation that could contribute to the constructive interference, but due to their smaller power, we do not include them in the \avg{z} maps.

\subsubsection{Unwarped simulation}	

We perform the same analysis on the unwarped model in Fig.~\ref{fig:z_vzed_evolution_ws} (right) with an identical setup of cylindrical bins and time interval. The diagonal black lines are the most prominent frequencies of the $m=1$ and $m=2$ bending waves taken from Figs.~\ref{fig:739HF_hc_psp} (bottom right) and~\ref{fig:739HF_hc_bubble} (right) for the $10-12\Gyr$ time interval. The values of the frequencies are indicated in the legend at the top of Fig.~\ref{fig:z_vzed_evolution_ws}.
    
The distributions of \avg{z} for the unwarped simulation show no dominant signal but a superposition of multiplicities with amplitudes that are weaker by a factor of 5 than in the warped model (note the different colour-scale). The most recognisable signals are a prograde $m=2$ ($t=10.5\Gyr$), a retrograde $m=1$ ($11.0\leq t /\Gyr\leq11.5\Gyr$), and an $m=0$ ($t=10.25\Gyr$) signal. From Figs.~\ref{fig:739HF_hc_psp} and \ref{fig:739HF_hc_bubble} we determine the pattern speeds of the most prominent bending waves and overlay them on top of the \avg{z} distribution. The bending waves with the most power are the retrograde $m=1$ (solid) and prograde $m=2$ (dotted) with $\Omega_{p}\approx-13\kms\kpc^{-1}$ and $\Omega_{p}\approx12\kms\kpc^{-1}$, respectively. These bending waves seemingly match the slopes of the $m=1$ and $m=2$ signals in the \avg{z} distribution at all radii, e.g\ at $t=11.0 \Gyr$ and $t=10.2 \Gyr$, respectively. We observe additional bending waves in Figs.~\ref{fig:739HF_hc_psp} and \ref{fig:739HF_hc_bubble} that have less power, \eg\ the prograde $m=1$ wave with $\Omega_{p}\approx24\kms\kpc^{-1}$. When overlaying this pattern speed on top of the \avg{z} distribution we observe some coincidences with the slopes of ripples. We again speculate that the visible signals in the \avg{z} distribution are caused  by constructive interference.

\subsubsection{Bending waves in the SN}
	\begin{figure}
		\includegraphics[width=1.\linewidth]{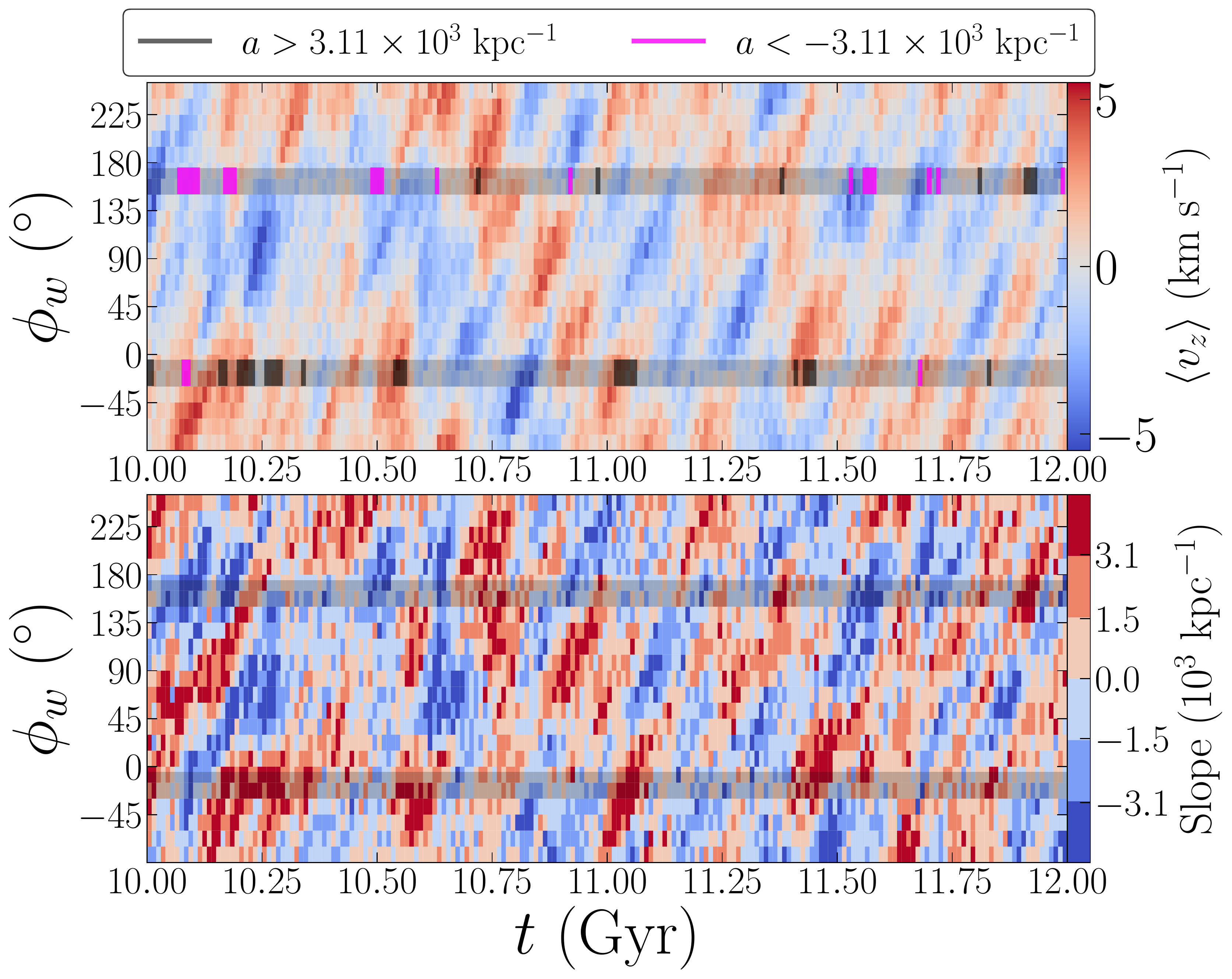}
		\caption{Evolution of the mean vertical velocity, \avg{v_z}, (top) and the slope of the $L_z$-$\avg{v_z}$ relation (bottom) in the radial interval $7.18\leq R /\kpc\leq9.18$ of the warped simulation. The horizontal shaded lines indicate a simulated SN ($\phi_w=-17.5\degrees$) and anti-SN ($\phi_w=162.5\degrees$) samples in which the slope of the $L_z$-$\avg{v_z}$ is measured at each time interval. The vertical black lines indicate where the slope of the $L_z$-$\avg{v_z}$ relation is equal or exceeds the slope as measured by \protect\citetalias{wrinkle} and \protect\citetalias{huang}. In the top panel, the vertical magenta lines indicate where the slope of the $L_z$-$\avg{v_z}$ relation is equal or less than the negative of this observed value. We observe that large positive and negative slopes are correlated with the peaks and troughs of bending waves, respectively.
		} 
		\label{fig:zvzed_slope}
	\end{figure}

After uncovering the complex bending signatures of the warped model in Fig.~\ref{fig:z_vzed_evolution_ws}, we test if the peaks of the bending waves correlate with large slope values examined in Section~\ref{subsec:sch}.  Higher positive slope values are likely to manifest when a \avg{v_z} peak (red) of a bending wave is passing through the SN. In Fig.~\ref{fig:zvzed_slope} we use the same sectoral bins implemented in Fig.~\ref{fig:z_vzed_evolution_ws} to measure the evolution of \avg{z} (top) and the slope of the $L_z$-$\avg{v_z}$ relation (bottom) in each azimuthal bin at $7.18\leq R /\kpc\leq9.18$ (Solar annulus). We indicate the SN and anti-SN samples with a shaded grey line centred on $\phi_w=-17.5\degrees$ and $\phi_w=162.5\degrees$ \citep{chen}, respectively, with an azimuthal range of $\delta \phi_w = 25\degrees$. At each time interval we fit Eq.~\ref{eq:lin} to the $L_z$-$\avg{v_z}$ distribution in each azimuthal bin; when the slope of the fit is greater than the one measured by \protect\citetalias{wrinkle} and \protect\citetalias{huang}, the time interval is indicated with a black vertical line.

The slope in the $L_z$-$\avg{v_z}$ of the SN appears to exceed the \protect\citetalias{wrinkle} and \protect\citetalias{huang} values when the SN has a peak in \avg{v_z}. The slope distribution in the bottom panel demonstrates that the peaks correlate with $a\gtrsim 1 \times10^3 \kpc^{-1}$. No time interval with a trough (blue) in the SN sample has a slope that exceeds the observed value, and are mostly negative. The anti-SN has fewer positive slopes and more regions with negative slopes that reach $a\lesssim -3.05 \times10^3 \kpc^{-1}$ (vertical magenta lines) with a similar frequency as the \protect\citetalias{wrinkle} and \protect\citetalias{huang} values is reached in the SN sample. Further analysis of earlier times indicate that this inversion is not present in both SN and anti-SN samples, but rather reach the \protect\citetalias{wrinkle} and \protect\citetalias{huang} and their negative values at similar rates. The inversion at $10\leq t / \Gyr \leq 12$ is likely the result of gas flux variations between the North ($z>0$) and South ($z<0$) sides of the warp and not due to the location of our samples.

The slope evolution demonstrates that regions of $a\gtrsim 3.05 \times10^3 \kpc^{-1}$ and $a\lesssim -3.05 \times10^3 \kpc^{-1}$ propagate through the entire Solar annulus in a retrograde fashion, mirroring the dominant $m=1$ signal found in Fig.~\ref{fig:z_vzed_evolution_ws}. Additionally, the ripples in the \avg{z} (Fig.~\ref{fig:z_vzed_evolution_ws}) and \avg{v_z} distributions (that we interpret as the result of constructive interference) are also present in the slope evolution. The results of this analysis can be interpreted as a direct link between the high slope values observed in \protect\citetalias{wrinkle} and \protect\citetalias{huang} and the bending waves manifesting from the warp's perturbation.

\subsection{The effect of stellar ages}
	\begin{figure*}
		\includegraphics[width=1\linewidth]{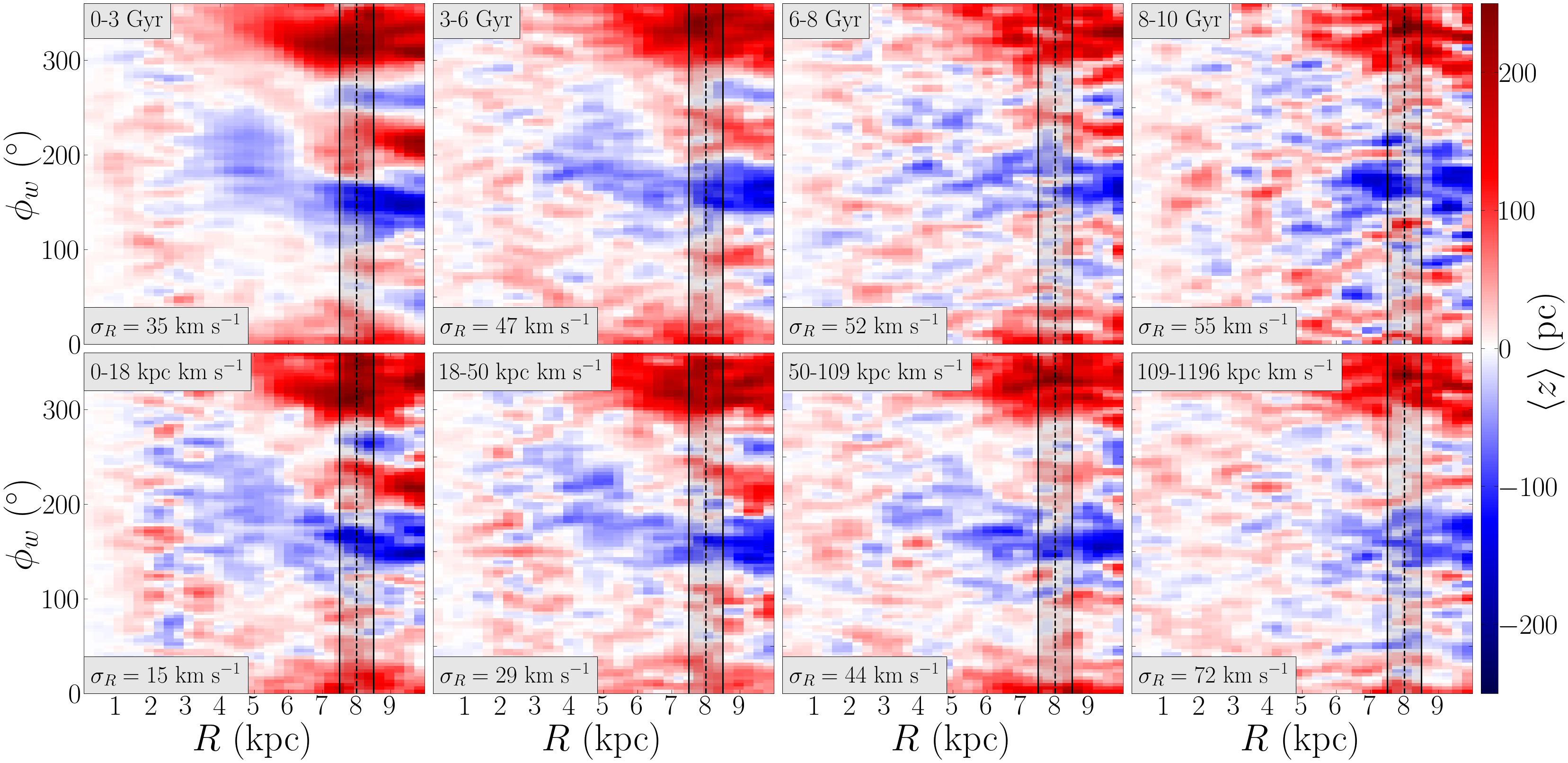}
		
		\caption{Cylindrical distributions of $\avg{z}$ for populations in different age (top) and radial action, $J_R$, (bottom) ranges in the warped model at $10.8 \Gyr$ (age and $J_R$ ranges annotated in the top left of each panel). The bottom left of each panel shows the radial velocity dispersion, $\sigma_R$, in the Solar annulus sample of each age and $J_R$ cuts (shaded region). The sense of rotation is towards increasing $\phi_w$, where $\phi_w=0\degrees$ represents the ascending node of the warp's LON. A bending wave is visible in all populations but is strongest, and most clearly defined, in the youngest and low-$J_R$ populations. A Gaussian filter has been applied to the colour distribution in each panel with a standard deviation of the Gaussian kernel set to $\sigma=1$ pixel $=570\times 570 \pc$.} 
		\label{fig:z_vz_age}
	\end{figure*}

The dispersion relation of Eq.~\eqref{eq:HT_DR} applies to WKB waves in kinematically cold discs, \ie\ in the absence of in-plane velocity dispersion.
In real discs, scattering at inner and outer Lindblad resonances and with giant molecular clouds increase the orbital eccentricity of stars, making populations kinematically hotter over time \citep[see \eg\ ][for a review]{Sellwood14}. \cite{Malkus_Thayer_1966} derived the dispersion relation of small-amplitude bending waves $h(x,t)=He^{i(kx-\omega t)}$ propagating in an infinite, thin slab of stars
	\begin{equation}
        \label{eq:Toomre_disp}
        \omega^2 =  2\upi G\Sigma(R)|k| - \sigma_x^{2} k^2,
    \end{equation}
where $\sigma_x$ is the in-plane velocity dispersion in the $x$-direction and $\Sigma$ is the vertically-integrated surface density of the slab. Similar to Eq.~\eqref{eq:HT_DR}, the term $2\upi G\Sigma(R)|k|$ represents the gravitational restoring force brought on by the bend, while $\sigma_x^{2} k^2$ is a destabilising inertial term. This dispersion relation demonstrates that disturbances with shorter wavelengths (large $k$) are unstable as they outweigh the restoring force, which translates to an exponential growth of the $h(x,t)$ distortion.

The assumption of an infinite, thin slab of stars cannot be applied to real galaxies, but it provides a useful estimate of how the propagation of bending waves is affected by the in-plane velocity dispersion. For the purpose of this analysis, we substitute the $\sigma_x$ in Eq.~\eqref{eq:Toomre_disp} with the in-plane radial velocity dispersion, $\sigma_R$. For a disc with a given $\sigma_R$, bending waves are able to propagate provided $k$ is smaller than the cutoff value which ensures that the right hand side of Eqn.~\ref{eq:Toomre_disp} remains positive. As $\sigma_R$ rises this critical $k$ needs to decrease. While this holds for discs with different $\sigma_R$, we might suspect that, within a given disc, kinematically hotter populations will not be able to support short wavelength bending waves. For any wavepacket, which is constructed by the superposition of sinusoidal waves of varying wavelength, the shorter wavelengths may only be supported by the kinematically coolest populations. As such the wavepacket might be expected to be sharper in cooler populations, and more gently varying in the hotter populations. As $\sigma_R$ rises with stellar age in a stellar disc, we test whether the bending waves are sharper in younger populations. 

The top row of Fig.~\ref{fig:z_vz_age} presents the distributions of \avg{z} at $t=10.8 \Gyr$ for populations separated by stellar age, in four equally populated bins. The distributions are presented for stars formed in the main disc only, in order to avoid warp stars that can take up to $\sim 6 \Gyr$ to fully settle and phase mix into the disc \citep{Khachaturyants+21}.
Besides the overall $m=1$ bend, we observe strong bending waves, in the \avg{z} distributions (coherent red and blue structures). These bending waves reach as far inside the disc as $R=4 \kpc$ in the youngest population, with amplitudes of $\sim 100\pc$. More importantly, the bending waves, while strongest in the young populations, can be recognised  in all populations. The bending waves in the youngest population are also the sharpest ones, whereas the waves in the older populations become less sharp at short wavelengths. Thus, in the old populations, Fig.~\ref{fig:z_vz_age} reveals a coherent large-scale signal but not so much the small scale patterns present in the young populations.

In order to aid in comparing with observational data, for which stellar ages have high uncertainties \citep{sanders_das}, we split the stars formed in the main disc by their radial actions, $J_R$. The radial action of a star characterises the extent of radial oscillations of a star's orbit and is thus a proxy for the in-plane velocity dispersion. The bottom row of Fig.~\ref{fig:z_vz_age} presents the distributions of \avg{z}\ separated by $J_R$ in bins containing an equal number of stars. Bending waves become less sharp with increasing values of $J_R$, similar to the age cuts in Fig.~\ref{fig:z_vz_age}.
	
\begin{figure*}
	\includegraphics[width=1\linewidth]{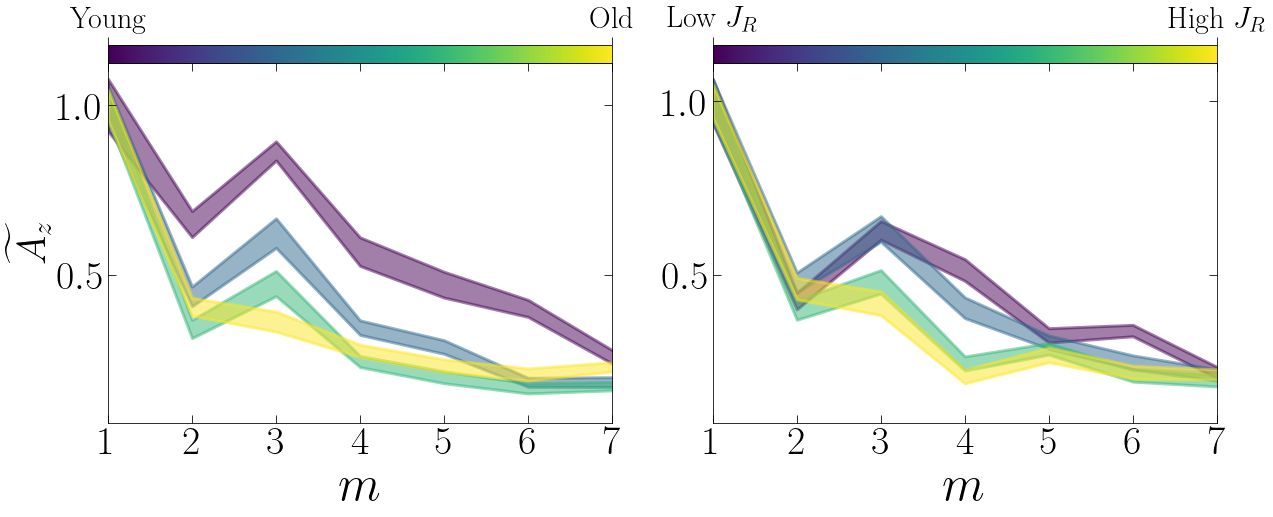}
	\caption{Relationship between normalised vertical displacement amplitudes, $\widetilde{A_z}$ (see Eq.~\ref{eq:gamma_m}), and their respective Fourier multiplicities, $m$. The Fourier amplitudes are calculated for the different stellar populations presented in the shaded region of Fig.~\ref{fig:z_vz_age}. In order to reduce the effect of noise, the amplitudes are calculated for 50 sequential snapshots ($\Delta t = 500\Myr$), averaged over the entire time interval, and finally normalised by $\widetilde{A_z}(m=1)$. The shaded region shows the standard deviation of $\widetilde{A_z}$.
	}
	\label{fig:az_40}
\end{figure*}
	
As the age and $J_R$ bins are equally populated, we expect similar levels of Poisson noise and, therefore, equally sharp bending waves if their sharpness was only noise limited; however, there is a clear difference between these populations. This implies that the kinematically hottest populations are unable to support the shortest wavelengths.

For a quantitative demonstration that bending waves are sharper in kinematically cooler populations, we note that short wavelengths correspond to larger Fourier $m$ terms compared to long wavelengths. Thus bending wave power should be concentrated in lower Fourier $m$ components in hot populations compared with cool ones.
We analyse the Fourier amplitudes of the vertical displacement for the shaded populations in Fig.~\ref{fig:z_vz_age}. We calculate the amplitudes using Eqn.~\ref{eq:gamma_m} for $m\in[1,7]$ in 50 sequential snapshots, which corresponds to the time interval $10.5\leq t /\Gyr\leq 11.0$. The centre-left panel of Fig.~\ref{fig:z_vzed_evolution_ws} shows that a clear $m=1$ bending wave is present in this region at this time. We avoid larger $m$ values since these mostly capture noise in the vertical distribution. In order to inter-compare the different populations, the amplitudes are averaged over the $500 \Myr$ time interval and then normalised by their $m=1$ amplitudes. In this way the difference in the overall bending wave strength in the different populations is factored out. The variation of the normalised vertical displacement amplitudes, $\widetilde{A_z}$, with multiplicity is presented in Fig.~\ref{fig:az_40}, with the shaded regions representing the standard deviation of $\widetilde{A_z}$. The amplitude of the vertical displacement decreases more rapidly with $m$ for increasing age and $J_R$ populations, indicating that the bending wave power is restricted to large wavelengths in the kinematically hot populations and confirming that the bending waves are sharper in the cool, young populations and smoother in the hotter, old ones. This agrees with the visual impression of the waves given by Fig.~\ref{fig:z_vz_age}. We have repeated this analysis at wide range of radii and times and have confirmed that in the vast majority of cases the power is restricted to smaller $m$ in hot populations compared with cooler ones.

\section{Summary}
\label{sec:Conclusions}

In this paper we have analysed the bending waves that appear in simulations with no recent interactions. One of these simulations develops a warp via continuous misaligned gas accretion. We demonstrated that both simulations exhibit bending waves. However, the presence of the warp produced significantly stronger bending waves, as well as more substantial power in prograde bending waves.
\begin{enumerate}
\setlength\itemsep{1em}
\item The warped model produces both retrograde and prograde bending waves; the latter would normally decay rapidly, but are continuously re-excited throughout the model's evolution. Bending waves in the unwarped model are significantly weaker in amplitude when compared to the warped model. The prograde waves are damped with time while the retrograde waves are long-lived. The pattern speeds of the bending waves in both simulations are consistent with theoretically expected "forbidden regions" for bending waves.
\item The irregular gas inflow in the warped simulation is the source of the perturbations exciting the stronger bending waves, with the flux of cold gas correlated to the strength of the bending waves of the disc. The flux varies substantially over time and on a wide range of frequencies. Cross-correlation analysis between the flux and vertical power at two annuli revealed a lag indicative of an inward-propagating bending wave. The group velocity estimated from the time lag is very similar to the radial group velocity obtained from the WKB approximation.
\item In the simulated Solar Neighbourhood (SN) sample of the warped model, the slope of the $L_z$-\avg{v_z} relation reaches and exceeds the values observed in the SN \citep{wrinkle}. The slope shows a wave-like dependence on azimuth with the wave propagating in a prograde direction. Since the warp is fixed in our analysis, this indicates that the slope is a result of propagating bending waves and not a direct imprint of the warp. In the unwarped simulation, the smaller bending wave amplitudes result in weaker slopes of the $L_z$-\avg{v_z} relation; however, the slopes still reach $\sim60\%$ of the \cite{wrinkle} values.
\item In the warped simulation, bending waves are present in stellar populations of all ages. The strongest and sharpest waves in $\avg{z}$ are in the youngest populations, while older ones are less sharp. The same trends are observed when splitting populations by the radial action.
\item The azimuthal \avg{z} distribution in the warped model exhibits a strong, retrograde, $m=1$ bend, extending inwards to at least $R = 5\kpc$. We also find localised prograde bending waves, which we term ripples. The overlaid frequencies from the spectral analysis match the slopes of $m=2$ and $m=1$ prograde ripples, suggesting they are the result of constructive interference between the prograde waves and the long-lived retrograde $m=1$ bending wave.
\end{enumerate}

In conclusion, our results demonstrate the role of misaligned gas accretion as a continuous source of vertical perturbations and the excitation of bending waves in galactic discs. This process may also contribute to the vertical perturbations and out-of-equilibrium features observed in the Solar Neighbourhood.

	\section*{Acknowledgements}
	
	V.P.D., L.B.S. and T.K. are supported by STFC Consolidated grant \#ST/R000786/1. We thank the referee of this paper, James Binney, for comments that helped improve the paper. The simulations in this paper were run at the DiRAC Shared Memory Processing system at the University of Cambridge, operated by the COSMOS Project at the Department of Applied Mathematics and Theoretical Physics on behalf of the STFC DiRAC HPC Facility (www.dirac.ac.uk). This equipment was funded by BIS National E-infrastructure capital grant ST/J005673/1, STFC capital grant ST/H008586/1 and STFC DiRAC Operations grant ST/K00333X/1. DiRAC is part of the National E-Infrastructure. Analysis was carried out on Stardynamics, a 64 core machine which was funded from Newton Advanced Fellowship NA150272 awarded by the Royal Society and the Newton Fund. All of the simulation processing was performed with the \textsc{Python} library \textsc{Pynbody} \citep{pynbody}. The authors thank Rok Ro\v skar for providing us with his code for producing spectrograms and Sarah Loebman for suggesting using this code for the bending wave spectrum.


\section*{Data availability}

The simulation dataset used here can be shared for limited use on request to V.P.D. (vpdebattista@gmail.com).

	\bibliographystyle{mnras}
	\bibliography{references} 

\begin{thebibliography}{}
\makeatletter
\relax
\def\mn@urlcharsother{\let\do\@makeother \do\$\do\&\do\#\do\^\do\_\do\%\do\~}
\def\mn@doi{\begingroup\mn@urlcharsother \@ifnextchar [ {\mn@doi@}
  {\mn@doi@[]}}
\def\mn@doi@[#1]#2{\def\@tempa{#1}\ifx\@tempa\@empty \href
  {http://dx.doi.org/#2} {doi:#2}\else \href {http://dx.doi.org/#2} {#1}\fi
  \endgroup}
\def\mn@eprint#1#2{\mn@eprint@#1:#2::\@nil}
\def\mn@eprint@arXiv#1{\href {http://arxiv.org/abs/#1} {{\tt arXiv:#1}}}
\def\mn@eprint@dblp#1{\href {http://dblp.uni-trier.de/rec/bibtex/#1.xml}
  {dblp:#1}}
\def\mn@eprint@#1:#2:#3:#4\@nil{\def\@tempa {#1}\def\@tempb {#2}\def\@tempc
  {#3}\ifx \@tempc \@empty \let \@tempc \@tempb \let \@tempb \@tempa \fi \ifx
  \@tempb \@empty \def\@tempb {arXiv}\fi \@ifundefined
  {mn@eprint@\@tempb}{\@tempb:\@tempc}{\expandafter \expandafter \csname
  mn@eprint@\@tempb\endcsname \expandafter{\@tempc}}}

\bibitem[\protect\citeauthoryear{{Agertz}, {Teyssier}  \& {Moore}}{{Agertz}
  et~al.}{2009}]{Agertz+2009}
{Agertz} O.,  {Teyssier} R.,   {Moore} B.,  2009, \mn@doi [\mnras]
  {10.1111/j.1745-3933.2009.00685.x}, \href
  {https://ui.adsabs.harvard.edu/abs/2009MNRAS.397L..64A} {397, L64}

\bibitem[\protect\citeauthoryear{Antoja et~al.,}{Antoja et~al.}{2018}]{antoja}
Antoja T.,  et~al., 2018, \mn@doi [Nature] {10.1038/s41586-018-0510-7}, 561,
  360

\bibitem[\protect\citeauthoryear{{Aumer} \& {White}}{{Aumer} \&
  {White}}{2013}]{AumerWhite2013}
{Aumer} M.,  {White} S. D.~M.,  2013, \mn@doi [\mnras] {10.1093/mnras/sts083},
  \href {https://ui.adsabs.harvard.edu/abs/2013MNRAS.428.1055A} {428, 1055}

\bibitem[\protect\citeauthoryear{{Aumer}, {White}, {Naab}  \&
  {Scannapieco}}{{Aumer} et~al.}{2013}]{aumeraccr}
{Aumer} M.,  {White} S. D.~M.,  {Naab} T.,   {Scannapieco} C.,  2013, \mn@doi
  [\mnras] {10.1093/mnras/stt1230}, \href
  {https://ui.adsabs.harvard.edu/abs/2013MNRAS.434.3142A} {434, 3142}

\bibitem[\protect\citeauthoryear{{Bennett} \& {Bovy}}{{Bennett} \&
  {Bovy}}{2021}]{Bennett+21}
{Bennett} M.,  {Bovy} J.,  2021, \mn@doi [\mnras] {10.1093/mnras/stab524},
  \href {https://ui.adsabs.harvard.edu/abs/2021MNRAS.503..376B} {503, 376}

\bibitem[\protect\citeauthoryear{{Binney} \& {May}}{{Binney} \&
  {May}}{1986}]{binney_may86}
{Binney} J.,  {May} A.,  1986, \mn@doi [\mnras] {10.1093/mnras/218.4.743},
  \href {https://ui.adsabs.harvard.edu/abs/1986MNRAS.218..743B} {218, 743}

\bibitem[\protect\citeauthoryear{{Binney} \& {Sch{\"o}nrich}}{{Binney} \&
  {Sch{\"o}nrich}}{2018}]{binney}
{Binney} J.,  {Sch{\"o}nrich} R.,  2018, \mn@doi [\mnras]
  {10.1093/mnras/sty2378}, \href
  {https://ui.adsabs.harvard.edu/abs/2018MNRAS.481.1501B} {481, 1501}

\bibitem[\protect\citeauthoryear{{Binney} \& {Tremaine}}{{Binney} \&
  {Tremaine}}{2008}]{binney_tremaine_08}
{Binney} J.,  {Tremaine} S.,  2008, {Galactic Dynamics: Second Edition}.
Princeton Univ. Press

\bibitem[\protect\citeauthoryear{{Binney}, {Jiang}  \& {Dutta}}{{Binney}
  et~al.}{1998}]{Binney_1998}
{Binney} J.,  {Jiang} I.-G.,   {Dutta} S.,  1998, \mn@doi [MNRAS]
  {10.1046/j.1365-8711.1998.01595.x}, \href
  {https://ui.adsabs.harvard.edu/abs/1998MNRAS.297.1237B} {297, 1237}

\bibitem[\protect\citeauthoryear{{Bland-Hawthorn} \&
  {Tepper-Garc{\'\i}a}}{{Bland-Hawthorn} \&
  {Tepper-Garc{\'\i}a}}{2021}]{Hawthorn+21}
{Bland-Hawthorn} J.,  {Tepper-Garc{\'\i}a} T.,  2021, \mn@doi [\mnras]
  {10.1093/mnras/stab704}, \href
  {https://ui.adsabs.harvard.edu/abs/2021MNRAS.504.3168B} {504, 3168}

\bibitem[\protect\citeauthoryear{{Bland-Hawthorn} et~al.,}{{Bland-Hawthorn}
  et~al.}{2019}]{galah}
{Bland-Hawthorn} J.,  et~al., 2019, \mn@doi [\mnras] {10.1093/mnras/stz217},
  \href {https://ui.adsabs.harvard.edu/abs/2019MNRAS.486.1167B} {486, 1167}

\bibitem[\protect\citeauthoryear{{Briggs}}{{Briggs}}{1990}]{briggs}
{Briggs} F.~H.,  1990, \mn@doi [\apj] {10.1086/168512}, \href
  {http://adsabs.harvard.edu/abs/1990ApJ...352...15B} {352, 15}

\bibitem[\protect\citeauthoryear{{Bullock}, {Dekel}, {Kolatt}, {Kravtsov},
  {Klypin}, {Porciani}  \& {Primack}}{{Bullock} et~al.}{2001}]{Bullock+2001}
{Bullock} J.~S.,  {Dekel} A.,  {Kolatt} T.~S.,  {Kravtsov} A.~V.,  {Klypin}
  A.~A.,  {Porciani} C.,   {Primack} J.~R.,  2001, \mn@doi [\apj]
  {10.1086/321477}, \href
  {https://ui.adsabs.harvard.edu/abs/2001ApJ...555..240B} {555, 240}

\bibitem[\protect\citeauthoryear{{Carlin} et~al.,}{{Carlin}
  et~al.}{2013}]{Carlin+13}
{Carlin} J.~L.,  et~al., 2013, \mn@doi [\apjl] {10.1088/2041-8205/777/1/L5},
  \href {https://ui.adsabs.harvard.edu/abs/2013ApJ...777L...5C} {777, L5}

\bibitem[\protect\citeauthoryear{{Chen}, {Wang}, {Deng}, {de Grijs}  \&
  {Yang}}{{Chen} et~al.}{2018}]{wise}
{Chen} X.,  {Wang} S.,  {Deng} L.,  {de Grijs} R.,   {Yang} M.,  2018, \mn@doi
  [\apjs] {10.3847/1538-4365/aad32b}, \href
  {https://ui.adsabs.harvard.edu/abs/2018ApJS..237...28C} {237, 28}

\bibitem[\protect\citeauthoryear{{Chen}, {Wang}, {Deng}, {de Grijs}, {Liu}  \&
  {Tian}}{{Chen} et~al.}{2019}]{chen}
{Chen} X.,  {Wang} S.,  {Deng} L.,  {de Grijs} R.,  {Liu} C.,   {Tian} H.,
  2019, \mn@doi [Nature Astronomy] {10.1038/s41550-018-0686-7}, \href
  {https://ui.adsabs.harvard.edu/abs/2019NatAs...3..320C} {3, 320}

\bibitem[\protect\citeauthoryear{{Chequers} \& {Widrow}}{{Chequers} \&
  {Widrow}}{2017}]{chequers+17}
{Chequers} M.~H.,  {Widrow} L.~M.,  2017, \mn@doi [\mnras]
  {10.1093/mnras/stx2165}, \href
  {https://ui.adsabs.harvard.edu/abs/2017MNRAS.472.2751C} {472, 2751}

\bibitem[\protect\citeauthoryear{{Chequers}, {Widrow}  \& {Darling}}{{Chequers}
  et~al.}{2018}]{cheque}
{Chequers} M.~H.,  {Widrow} L.~M.,   {Darling} K.,  2018, \mn@doi [\mnras]
  {10.1093/mnras/sty2114}, \href
  {https://ui.adsabs.harvard.edu/abs/2018MNRAS.480.4244C} {480, 4244}

\bibitem[\protect\citeauthoryear{{Darling} \& {Widrow}}{{Darling} \&
  {Widrow}}{2019}]{darling}
{Darling} K.,  {Widrow} L.~M.,  2019, \mn@doi [\mnras] {10.1093/mnras/sty3508},
  \href {https://ui.adsabs.harvard.edu/abs/2019MNRAS.484.1050D} {484, 1050}

\bibitem[\protect\citeauthoryear{{Debattista}, {van den Bosch}, {Ro{\v{s}}kar},
  {Quinn}, {Moore}  \& {Cole}}{{Debattista} et~al.}{2015}]{vpd2015}
{Debattista} V.~P.,  {van den Bosch} F.~C.,  {Ro{\v{s}}kar} R.,  {Quinn} T.,
  {Moore} B.,   {Cole} D.~R.,  2015, \mn@doi [\mnras] {10.1093/mnras/stv1563},
  \href {https://ui.adsabs.harvard.edu/abs/2015MNRAS.452.4094D} {452, 4094}

\bibitem[\protect\citeauthoryear{{Debattista}, {Gonzalez}, {Sand erson},
  {El-Badry}, {Garrison-Kimmel}, {Wetzel}, {Faucher-Gigu{\`e}re}  \&
  {Hopkins}}{{Debattista} et~al.}{2019}]{vpd2019}
{Debattista} V.~P.,  {Gonzalez} O.~A.,  {Sand erson} R.~E.,  {El-Badry} K.,
  {Garrison-Kimmel} S.,  {Wetzel} A.,  {Faucher-Gigu{\`e}re} C.-A.,   {Hopkins}
  P.~F.,  2019, \mn@doi [\mnras] {10.1093/mnras/stz746}, \href
  {https://ui.adsabs.harvard.edu/abs/2019MNRAS.485.5073D} {485, 5073}

\bibitem[\protect\citeauthoryear{{Dehnen}}{{Dehnen}}{1998}]{sgr_dehnen}
{Dehnen} W.,  1998, \mn@doi [\aj] {10.1086/300364}, \href
  {https://ui.adsabs.harvard.edu/abs/1998AJ....115.2384D} {115, 2384}

\bibitem[\protect\citeauthoryear{{Dekel} \& {Shlosman}}{{Dekel} \&
  {Shlosman}}{1983}]{Dekel&Shlosman83}
{Dekel} A.,  {Shlosman} I.,  1983, in {Athanassoula} E.,  ed.,  Vol. 100,
  Internal Kinematics and Dynamics of Galaxies. p.~187

\bibitem[\protect\citeauthoryear{{Earp}, {Debattista}, {Macci{\`o}}  \&
  {Cole}}{{Earp} et~al.}{2017}]{earp+17}
{Earp} S. W.~F.,  {Debattista} V.~P.,  {Macci{\`o}} A.~V.,   {Cole} D.~R.,
  2017, \mn@doi [\mnras] {10.1093/mnras/stx1143}, \href
  {https://ui.adsabs.harvard.edu/abs/2017MNRAS.469.4095E} {469, 4095}

\bibitem[\protect\citeauthoryear{{Earp}, {Debattista}, {Macci{\`o}}, {Wang},
  {Buck}  \& {Khachaturyants}}{{Earp} et~al.}{2019}]{earp+19}
{Earp} S. W.~F.,  {Debattista} V.~P.,  {Macci{\`o}} A.~V.,  {Wang} L.,  {Buck}
  T.,   {Khachaturyants} T.,  2019, \mn@doi [\mnras] {10.1093/mnras/stz2109},
  \href {https://ui.adsabs.harvard.edu/abs/2019MNRAS.488.5728E} {488, 5728}

\bibitem[\protect\citeauthoryear{{Efremov}, {Ivanov}  \& {Nikolov}}{{Efremov}
  et~al.}{1981}]{efremov}
{Efremov} Y.~N.,  {Ivanov} G.~R.,   {Nikolov} N.~S.,  1981, \mn@doi [\apss]
  {10.1007/BF00648652}, \href
  {https://ui.adsabs.harvard.edu/abs/1981Ap&SS..75..407E} {75, 407}

\bibitem[\protect\citeauthoryear{{Faure}, {Siebert}  \& {Famaey}}{{Faure}
  et~al.}{2014}]{Faure+14}
{Faure} C.,  {Siebert} A.,   {Famaey} B.,  2014, \mn@doi [\mnras]
  {10.1093/mnras/stu428}, \href
  {https://ui.adsabs.harvard.edu/abs/2014MNRAS.440.2564F} {440, 2564}

\bibitem[\protect\citeauthoryear{Feldmann \& Spolyar}{Feldmann \&
  Spolyar}{2014}]{feldman}
Feldmann R.,  Spolyar D.,  2014, \mn@doi [MNRAS] {10.1093/mnras/stu2147}, 446,
  1000

\bibitem[\protect\citeauthoryear{{Fiteni}, {Caruana}, {Amarante}, {Debattista}
  \& {Beraldo e Silva}}{{Fiteni} et~al.}{2021}]{karl21+}
{Fiteni} K.,  {Caruana} J.,  {Amarante} J. A.~S.,  {Debattista} V.~P.,
  {Beraldo e Silva} L.,  2021, \mn@doi [\mnras] {10.1093/mnras/stab619}, \href
  {https://ui.adsabs.harvard.edu/abs/2021MNRAS.503.1418F} {503, 1418}

\bibitem[\protect\citeauthoryear{{Foreman-Mackey}, {Hogg}, {Lang}  \&
  {Goodman}}{{Foreman-Mackey} et~al.}{2013}]{Foreman-Mackey+2019}
{Foreman-Mackey} D.,  {Hogg} D.~W.,  {Lang} D.,   {Goodman} J.,  2013, \mn@doi
  [PASP] {10.1086/670067}, \href
  {https://ui.adsabs.harvard.edu/abs/2013PASP..125..306F} {125, 306}

\bibitem[\protect\citeauthoryear{{Fox}, {Richter}, {Ashley}, {Heckman},
  {Lehner}, {Werk}, {Bordoloi}  \& {Peeples}}{{Fox}
  et~al.}{2019}]{gas_inflow_est}
{Fox} A.~J.,  {Richter} P.,  {Ashley} T.,  {Heckman} T.~M.,  {Lehner} N.,
  {Werk} J.~K.,  {Bordoloi} R.,   {Peeples} M.~S.,  2019, \mn@doi [\apj]
  {10.3847/1538-4357/ab40ad}, \href
  {https://ui.adsabs.harvard.edu/abs/2019ApJ...884...53F} {884, 53}

\bibitem[\protect\citeauthoryear{{Friske} \& {Sch{\"o}nrich}}{{Friske} \&
  {Sch{\"o}nrich}}{2019}]{friske+19}
{Friske} J. K.~S.,  {Sch{\"o}nrich} R.,  2019, \mn@doi [\mnras]
  {10.1093/mnras/stz2951}, \href
  {https://ui.adsabs.harvard.edu/abs/2019MNRAS.490.5414F} {490, 5414}

\bibitem[\protect\citeauthoryear{{Gaia Collaboration} et~al.,}{{Gaia
  Collaboration} et~al.}{2016a}]{gaia1}
{Gaia Collaboration} et~al., 2016a, \mn@doi [\aap]
  {10.1051/0004-6361/201629272}, \href
  {https://ui.adsabs.harvard.edu/abs/2016A&A...595A...1G} {595, A1}

\bibitem[\protect\citeauthoryear{{Gaia Collaboration} et~al.,}{{Gaia
  Collaboration} et~al.}{2016b}]{gaia}
{Gaia Collaboration} et~al., 2016b, \mn@doi [\aap]
  {10.1051/0004-6361/201629512}, \href
  {https://ui.adsabs.harvard.edu/abs/2016A&A...595A...2G} {595, A2}

\bibitem[\protect\citeauthoryear{{Gaia Collaboration} et~al.,}{{Gaia
  Collaboration} et~al.}{2018}]{dr2rv}
{Gaia Collaboration} et~al., 2018, \mn@doi [\aap]
  {10.1051/0004-6361/201832865}, \href
  {https://ui.adsabs.harvard.edu/abs/2018A&A...616A..11G} {616, A11}

\bibitem[\protect\citeauthoryear{{G{\'o}mez} et~al.,}{{G{\'o}mez}
  et~al.}{2012}]{Gomez+12}
{G{\'o}mez} F.~A.,  et~al., 2012, \mn@doi [\mnras]
  {10.1111/j.1365-2966.2012.21176.x}, \href
  {https://ui.adsabs.harvard.edu/abs/2012MNRAS.423.3727G} {423, 3727}

\bibitem[\protect\citeauthoryear{{G{\'o}mez}, {White}, {Grand }, {Marinacci},
  {Springel}  \& {Pakmor}}{{G{\'o}mez} et~al.}{2017}]{gomez}
{G{\'o}mez} F.~A.,  {White} S. D.~M.,  {Grand } R. J.~J.,  {Marinacci} F.,
  {Springel} V.,   {Pakmor} R.,  2017, \mn@doi [\mnras]
  {10.1093/mnras/stw2957}, \href
  {https://ui.adsabs.harvard.edu/abs/2017MNRAS.465.3446G} {465, 3446}

\bibitem[\protect\citeauthoryear{{Huang} et~al.,}{{Huang} et~al.}{2018}]{huang}
{Huang} Y.,  et~al., 2018, \mn@doi [\apj] {10.3847/1538-4357/aad285}, \href
  {https://ui.adsabs.harvard.edu/abs/2018ApJ...864..129H} {864, 129}

\bibitem[\protect\citeauthoryear{{Hunter} \& {Toomre}}{{Hunter} \&
  {Toomre}}{1969}]{hunter_toomre}
{Hunter} C.,  {Toomre} A.,  1969, \mn@doi [\apj] {10.1086/149908}, \href
  {https://ui.adsabs.harvard.edu/abs/1969ApJ...155..747H} {155, 747}

\bibitem[\protect\citeauthoryear{{Ibata} \& {Razoumov}}{{Ibata} \&
  {Razoumov}}{1998}]{sgr_ibata}
{Ibata} R.~A.,  {Razoumov} A.~O.,  1998, \aap, \href
  {https://ui.adsabs.harvard.edu/abs/1998A&A...336..130I} {336, 130}

\bibitem[\protect\citeauthoryear{{Jiang} \& {Binney}}{{Jiang} \&
  {Binney}}{1999}]{Jiang&Binney99}
{Jiang} I.-G.,  {Binney} J.,  1999, \mn@doi [\mnras]
  {10.1046/j.1365-8711.1999.02333.x}, \href
  {https://ui.adsabs.harvard.edu/abs/1999MNRAS.303L...7J} {303, L7}

\bibitem[\protect\citeauthoryear{{Kazantzidis}, {Zentner}, {Kravtsov},
  {Bullock}  \& {Debattista}}{{Kazantzidis} et~al.}{2009}]{Kazantzidis}
{Kazantzidis} S.,  {Zentner} A.~R.,  {Kravtsov} A.~V.,  {Bullock} J.~S.,
  {Debattista} V.~P.,  2009, \mn@doi [\apj] {10.1088/0004-637X/700/2/1896},
  \href {https://ui.adsabs.harvard.edu/abs/2009ApJ...700.1896K} {700, 1896}

\bibitem[\protect\citeauthoryear{{Kerr}}{{Kerr}}{1957}]{Kerr}
{Kerr} F.~J.,  1957, \mn@doi [\aj] {10.1086/107466}, \href
  {https://ui.adsabs.harvard.edu/abs/1957AJ.....62...93K} {62, 93}

\bibitem[\protect\citeauthoryear{{Khachaturyants}, {Beraldo e Silva}  \&
  {Debattista}}{{Khachaturyants} et~al.}{2021}]{Khachaturyants+21}
{Khachaturyants} T.,  {Beraldo e Silva} L.,   {Debattista} V.~P.,  2021,
  \mn@doi [\mnras] {10.1093/mnras/stab2653}, \href
  {https://ui.adsabs.harvard.edu/abs/2021MNRAS.tmp.2381K} {}

\bibitem[\protect\citeauthoryear{{Khoperskov}, {Di Matteo}, {Gerhard}, {Katz},
  {Haywood}, {Combes}, {Berczik}  \& {Gomez}}{{Khoperskov}
  et~al.}{2019}]{khoper}
{Khoperskov} S.,  {Di Matteo} P.,  {Gerhard} O.,  {Katz} D.,  {Haywood} M.,
  {Combes} F.,  {Berczik} P.,   {Gomez} A.,  2019, \mn@doi [\aap]
  {10.1051/0004-6361/201834707}, \href
  {https://ui.adsabs.harvard.edu/abs/2019A&A...622L...6K} {622, L6}

\bibitem[\protect\citeauthoryear{{Laporte}, {Minchev}, {Johnston}  \&
  {G{\'o}mez}}{{Laporte} et~al.}{2019}]{laporte2019}
{Laporte} C. F.~P.,  {Minchev} I.,  {Johnston} K.~V.,   {G{\'o}mez} F.~A.,
  2019, \mn@doi [\mnras] {10.1093/mnras/stz583}, \href
  {https://ui.adsabs.harvard.edu/abs/2019MNRAS.485.3134L} {485, 3134}

\bibitem[\protect\citeauthoryear{{Levine}, {Blitz}  \& {Heiles}}{{Levine}
  et~al.}{2006}]{2006ApJ...643..881L}
{Levine} E.~S.,  {Blitz} L.,   {Heiles} C.,  2006, \mn@doi [\apj]
  {10.1086/503091}, \href
  {https://ui.adsabs.harvard.edu/\#abs/2006ApJ...643..881L} {643, 881}

\bibitem[\protect\citeauthoryear{{Li} \& {Shen}}{{Li} \& {Shen}}{2020}]{juntai}
{Li} Z.-Y.,  {Shen} J.,  2020, \mn@doi [\apj] {10.3847/1538-4357/ab6b21}, \href
  {https://ui.adsabs.harvard.edu/abs/2020ApJ...890...85L} {890, 85}

\bibitem[\protect\citeauthoryear{{L{\'o}pez-Corredoira}, {Abedi}, {Garz{\'o}n}
  \& {Figueras}}{{L{\'o}pez-Corredoira} et~al.}{2014}]{lopez}
{L{\'o}pez-Corredoira} M.,  {Abedi} H.,  {Garz{\'o}n} F.,   {Figueras} F.,
  2014, \mn@doi [\aap] {10.1051/0004-6361/201424573}, \href
  {https://ui.adsabs.harvard.edu/abs/2014A&A...572A.101L} {572, A101}

\bibitem[\protect\citeauthoryear{{Merritt} \& {Sellwood}}{{Merritt} \&
  {Sellwood}}{1994}]{merritt_sellwood}
{Merritt} D.,  {Sellwood} J.~A.,  1994, \mn@doi [\apj] {10.1086/174005}, \href
  {https://ui.adsabs.harvard.edu/abs/1994ApJ...425..551M} {425, 551}

\bibitem[\protect\citeauthoryear{{Miller} \& {Scalo}}{{Miller} \&
  {Scalo}}{1979}]{MillerScalo1979}
{Miller} G.~E.,  {Scalo} J.~M.,  1979, \mn@doi [\apjs] {10.1086/190629}, \href
  {https://ui.adsabs.harvard.edu/abs/1979ApJS...41..513M} {41, 513}

\bibitem[\protect\citeauthoryear{{Navarro}, {Frenk}  \& {White}}{{Navarro}
  et~al.}{1996}]{NFW1996}
{Navarro} J.~F.,  {Frenk} C.~S.,   {White} S. D.~M.,  1996, \mn@doi [\apj]
  {10.1086/177173}, \href
  {https://ui.adsabs.harvard.edu/abs/1996ApJ...462..563N} {462, 563}

\bibitem[\protect\citeauthoryear{{Nelson} \& {Tremaine}}{{Nelson} \&
  {Tremaine}}{1995}]{nelson_tremaine_1995}
{Nelson} R.~W.,  {Tremaine} S.,  1995, \mn@doi [MNRAS]
  {10.1093/mnras/275.4.897}, \href
  {https://ui.adsabs.harvard.edu/abs/1995MNRAS.275..897N} {275, 897}

\bibitem[\protect\citeauthoryear{{Ostriker} \& {Binney}}{{Ostriker} \&
  {Binney}}{1989}]{ostriker_binney89}
{Ostriker} E.~C.,  {Binney} J.~J.,  1989, \mn@doi [\mnras]
  {10.1093/mnras/237.3.785}, \href
  {https://ui.adsabs.harvard.edu/abs/1989MNRAS.237..785O} {237, 785}

\bibitem[\protect\citeauthoryear{{Pontzen}, {Ro{\v s}kar}, {Stinson}, {Woods},
  {Reed}, {Coles}  \& {Quinn}}{{Pontzen} et~al.}{2013}]{pynbody}
{Pontzen} A.,  {Ro{\v s}kar} R.,  {Stinson} G.~S.,  {Woods} R.,  {Reed} D.~M.,
  {Coles} J.,   {Quinn} T.~R.,  2013, {pynbody: Astrophysics Simulation
  Analysis for Python}

\bibitem[\protect\citeauthoryear{{Raha}, {Sellwood}, {James}  \& {Kahn}}{{Raha}
  et~al.}{1991}]{raha+1991}
{Raha} N.,  {Sellwood} J.~A.,  {James} R.~A.,   {Kahn} F.~D.,  1991, \mn@doi
  [\nat] {10.1038/352411a0}, \href
  {https://ui.adsabs.harvard.edu/abs/1991Natur.352..411R} {352, 411}

\bibitem[\protect\citeauthoryear{{Reed}}{{Reed}}{1996}]{reed}
{Reed} B.~C.,  1996, \mn@doi [\aj] {10.1086/117826}, \href
  {https://ui.adsabs.harvard.edu/abs/1996AJ....111..804R} {111, 804}

\bibitem[\protect\citeauthoryear{{Ro{\v{s}}kar}, {Debattista}, {Brooks},
  {Quinn}, {Brook}, {Governato}, {Dalcanton}  \& {Wadsley}}{{Ro{\v{s}}kar}
  et~al.}{2010}]{roskar2010}
{Ro{\v{s}}kar} R.,  {Debattista} V.~P.,  {Brooks} A.~M.,  {Quinn} T.~R.,
  {Brook} C.~B.,  {Governato} F.,  {Dalcanton} J.~J.,   {Wadsley} J.,  2010,
  \mn@doi [\mnras] {10.1111/j.1365-2966.2010.17178.x}, \href
  {https://ui.adsabs.harvard.edu/abs/2010MNRAS.408..783R} {408, 783}

\bibitem[\protect\citeauthoryear{{Ro{\v{s}}kar}, {Debattista}, {Quinn}  \&
  {Wadsley}}{{Ro{\v{s}}kar} et~al.}{2012}]{Roskar+12}
{Ro{\v{s}}kar} R.,  {Debattista} V.~P.,  {Quinn} T.~R.,   {Wadsley} J.,  2012,
  \mn@doi [\mnras] {10.1111/j.1365-2966.2012.21860.x}, \href
  {https://ui.adsabs.harvard.edu/abs/2012MNRAS.426.2089R} {426, 2089}

\bibitem[\protect\citeauthoryear{{Sanders} \& {Das}}{{Sanders} \&
  {Das}}{2018}]{sanders_das}
{Sanders} J.~L.,  {Das} P.,  2018, \mn@doi [\mnras] {10.1093/mnras/sty2490},
  \href {https://ui.adsabs.harvard.edu/abs/2018MNRAS.481.4093S} {481, 4093}

\bibitem[\protect\citeauthoryear{{Sch{\"o}nrich} \& {Dehnen}}{{Sch{\"o}nrich}
  \& {Dehnen}}{2018}]{wrinkle}
{Sch{\"o}nrich} R.,  {Dehnen} W.,  2018, \mn@doi [\mnras]
  {10.1093/mnras/sty1256}, \href
  {https://ui.adsabs.harvard.edu/abs/2018MNRAS.478.3809S} {478, 3809}

\bibitem[\protect\citeauthoryear{{Sellwood}}{{Sellwood}}{1996}]{sellwood1996}
{Sellwood} J.~A.,  1996, \mn@doi [\apj] {10.1086/178185}, \href
  {https://ui.adsabs.harvard.edu/abs/1996ApJ...473..733S} {473, 733}

\bibitem[\protect\citeauthoryear{{Sellwood}}{{Sellwood}}{2014}]{Sellwood14}
{Sellwood} J.~A.,  2014, \mn@doi [Reviews of Modern Physics]
  {10.1103/RevModPhys.86.1}, \href
  {https://ui.adsabs.harvard.edu/abs/2014RvMP...86....1S} {86, 1}

\bibitem[\protect\citeauthoryear{{Sellwood} \& {Athanassoula}}{{Sellwood} \&
  {Athanassoula}}{1986}]{sellwood_athanassoula_1986}
{Sellwood} J.~A.,  {Athanassoula} E.,  1986, \mn@doi [\mnras]
  {10.1093/mnras/221.2.195}, \href
  {https://ui.adsabs.harvard.edu/abs/1986MNRAS.221..195S} {221, 195}

\bibitem[\protect\citeauthoryear{{Sellwood} \& {Merritt}}{{Sellwood} \&
  {Merritt}}{1994}]{sellwood}
{Sellwood} J.~A.,  {Merritt} D.,  1994, in American Astronomical Society
  Meeting Abstracts \#184. p. 36.01

\bibitem[\protect\citeauthoryear{{Sellwood}, {Nelson}  \&
  {Tremaine}}{{Sellwood} et~al.}{1998}]{sellwood1998}
{Sellwood} J.~A.,  {Nelson} R.~W.,   {Tremaine} S.,  1998, \mn@doi [\apj]
  {10.1086/306280}, \href
  {https://ui.adsabs.harvard.edu/abs/1998ApJ...506..590S} {506, 590}

\bibitem[\protect\citeauthoryear{{Shen}, {Wadsley}  \& {Stinson}}{{Shen}
  et~al.}{2010}]{Shen+2010}
{Shen} S.,  {Wadsley} J.,   {Stinson} G.,  2010, \mn@doi [\mnras]
  {10.1111/j.1365-2966.2010.17047.x}, \href
  {https://ui.adsabs.harvard.edu/abs/2010MNRAS.407.1581S} {407, 1581}

\bibitem[\protect\citeauthoryear{{Springel}}{{Springel}}{2010}]{Springel10}
{Springel} V.,  2010, \mn@doi [\araa] {10.1146/annurev-astro-081309-130914},
  \href {https://ui.adsabs.harvard.edu/abs/2010ARA&A..48..391S} {48, 391}

\bibitem[\protect\citeauthoryear{{Stevens}, {Lagos}, {Contreras}, {Croton},
  {Padilla}, {Schaller}, {Schaye}  \& {Theuns}}{{Stevens}
  et~al.}{2017}]{stevens}
{Stevens} A. R.~H.,  {Lagos} C. d.~P.,  {Contreras} S.,  {Croton} D.~J.,
  {Padilla} N.~D.,  {Schaller} M.,  {Schaye} J.,   {Theuns} T.,  2017, \mn@doi
  [\mnras] {10.1093/mnras/stx243}, \href
  {https://ui.adsabs.harvard.edu/abs/2017MNRAS.467.2066S} {467, 2066}

\bibitem[\protect\citeauthoryear{{Stinson}, {Seth}, {Katz}, {Wadsley},
  {Governato}  \& {Quinn}}{{Stinson} et~al.}{2006}]{Stinson+2006}
{Stinson} G.,  {Seth} A.,  {Katz} N.,  {Wadsley} J.,  {Governato} F.,   {Quinn}
  T.,  2006, \mn@doi [\mnras] {10.1111/j.1365-2966.2006.11097.x}, \href
  {https://ui.adsabs.harvard.edu/abs/2006MNRAS.373.1074S} {373, 1074}

\bibitem[\protect\citeauthoryear{Toomre}{Toomre}{1966}]{Malkus_Thayer_1966}
Toomre A.,  1966, in Geophysical Fluid Dynamics, notes on the 1966 Summer Study
  Program at the Woods Hole Oceanographic Institution, ref. no. 66-46,
  \mn@doi{10.1575/1912/2927}, \url {http://dx.doi.org/10.1575/1912/2927}

\bibitem[\protect\citeauthoryear{{Toomre}}{{Toomre}}{1983}]{Toomre83}
{Toomre} A.,  1983, in {Athanassoula} E.,  ed.,  Vol. 100, Internal Kinematics
  and Dynamics of Galaxies. pp 177--185

\bibitem[\protect\citeauthoryear{{Vasiliev}}{{Vasiliev}}{2019}]{agama}
{Vasiliev} E.,  2019, \mn@doi [\mnras] {10.1093/mnras/sty2672}, \href
  {https://ui.adsabs.harvard.edu/abs/2019MNRAS.482.1525V} {482, 1525}

\bibitem[\protect\citeauthoryear{{Velliscig} et~al.,}{{Velliscig}
  et~al.}{2015}]{velliscig}
{Velliscig} M.,  et~al., 2015, \mn@doi [\mnras] {10.1093/mnras/stv1690}, \href
  {https://ui.adsabs.harvard.edu/abs/2015MNRAS.453..721V} {453, 721}

\bibitem[\protect\citeauthoryear{{Wadsley}, {Stadel}  \& {Quinn}}{{Wadsley}
  et~al.}{2004}]{Wadsley+2004}
{Wadsley} J.~W.,  {Stadel} J.,   {Quinn} T.,  2004, \mn@doi [\na]
  {10.1016/j.newast.2003.08.004}, \href
  {https://ui.adsabs.harvard.edu/abs/2004NewA....9..137W} {9, 137}

\bibitem[\protect\citeauthoryear{{Weaver} \& {Williams}}{{Weaver} \&
  {Williams}}{1974}]{weaver_williams}
{Weaver} H.,  {Williams} D.~R.~W.,  1974, \aaps, \href
  {https://ui.adsabs.harvard.edu/abs/1974A&AS...17..251W} {17, 251}

\bibitem[\protect\citeauthoryear{{Werk} et~al.,}{{Werk}
  et~al.}{2019}]{Werk+2019}
{Werk} J.~K.,  et~al., 2019, \mn@doi [ApJ] {10.3847/1538-4357/ab54cf}, \href
  {https://ui.adsabs.harvard.edu/abs/2019ApJ...887...89W} {887, 89}

\bibitem[\protect\citeauthoryear{{Widrow}, {Gardner}, {Yanny}, {Dodelson}  \&
  {Chen}}{{Widrow} et~al.}{2012a}]{Widrow+12}
{Widrow} L.~M.,  {Gardner} S.,  {Yanny} B.,  {Dodelson} S.,   {Chen} H.-Y.,
  2012a, \mn@doi [\apjl] {10.1088/2041-8205/750/2/L41}, \href
  {https://ui.adsabs.harvard.edu/abs/2012ApJ...750L..41W} {750, L41}

\bibitem[\protect\citeauthoryear{Widrow, Gardner, Yanny, Dodelson  \&
  Chen}{Widrow et~al.}{2012b}]{widrow2012}
Widrow L.~M.,  Gardner S.,  Yanny B.,  Dodelson S.,   Chen H.-Y.,  2012b,
  \mn@doi [The Astrophysical Journal] {10.1088/2041-8205/750/2/l41}, 750, L41

\bibitem[\protect\citeauthoryear{{Williams} et~al.,}{{Williams}
  et~al.}{2013}]{Williams+13}
{Williams} M.~E.~K.,  et~al., 2013, \mn@doi [\mnras] {10.1093/mnras/stt1522},
  \href {https://ui.adsabs.harvard.edu/abs/2013MNRAS.436..101W} {436, 101}

\bibitem[\protect\citeauthoryear{{Yanny} \& {Gardner}}{{Yanny} \&
  {Gardner}}{2013}]{Yanny+13}
{Yanny} B.,  {Gardner} S.,  2013, \mn@doi [\apj] {10.1088/0004-637X/777/2/91},
  \href {https://ui.adsabs.harvard.edu/abs/2013ApJ...777...91Y} {777, 91}

\bibitem[\protect\citeauthoryear{{van den Bosch}, {Abel}, {Croft}, {Hernquist}
  \& {White}}{{van den Bosch} et~al.}{2002}]{vdb2002}
{van den Bosch} F.~C.,  {Abel} T.,  {Croft} R. A.~C.,  {Hernquist} L.,
  {White} S. D.~M.,  2002, \mn@doi [\apj] {10.1086/341619}, \href
  {https://ui.adsabs.harvard.edu/abs/2002ApJ...576...21V} {576, 21}

\makeatother
\end{thebibliography}


 \newcommand{\noop}[1]{}
	
	\bsp	
	
\end{document}